\documentclass[paper,notoc]{JHEP3}

\author{\footnote{KRYSTHAL Collaboration} L.A. Harland-Lang$^1$, V.A. Khoze$^{2,3}$, M.G. Ryskin$^{2,3}$, W.J. Stirling$^{1,2}$\\ 
  $^1$Cavendish Laboratory, University of Cambridge,
  J.J.\ Thomson Avenue, Cambridge, CB3 0HE, UK\\
  $^2$ Department of Physics and Institute for Particle Physics Phenomenology, University of Durham, Durham, DH1 3LE, UK\\
$^3$ Petersburg Nuclear Physics Institute, NRC Kurchatov Institute, Gatchina, \mbox{St. Petersburg}, 188300, Russia}

\title{The phenomenology of central exclusive production at hadron colliders}

\abstract{
Central exclusive production (CEP) processes in high--energy hadron--hadron collisions
provide an especially clean environment in which to measure the nature and quantum numbers (in particular, the spin and parity) of new resonance states. Encouraged by the broad agreement between experimental measurements and theoretical predictions based on the Durham approach, we perform a detailed phenomenological analysis of $\gamma\gamma$ and meson pair CEP final states, paying particular attention to the theoretical uncertainties in the predictions, including those from parton distribution functions, higher--order perturbative corrections, and non--perturbative and proton dissociation contributions. We present quantitative cross--section predictions for these CEP final states at the RHIC, Tevatron and LHC colliders.
}

\preprint{IPPP/12/12\\  DCPT/12/24\\ Cavendish-HEP-12/05}

\usepackage{cite}
\usepackage{subfigure}
\usepackage{multirow}
\usepackage{helvet}
\usepackage{amsmath}
\usepackage{amssymb}
\usepackage{setspace}
\usepackage{setspace}
\usepackage[dvips]{graphicx}
\usepackage{epsfig}
\def\lesim{ \;\raisebox{-.7ex}{$\stackrel{\textstyle <}{\sim}$}\; }
\def\be{\begin{equation}}
\def\ee{\end{equation}}

\begin{document}

\section{Introduction}

There has recently been a renewal of interest in studies of central exclusive production (CEP) processes in high--energy proton--(anti)proton collisions (see~\cite{Khoze:2001xm,Albrow:2010yb,Martin:2009ku,Albrow:2010zz} for reviews), both theoretically~\cite{HarlandLang:2009qe,Pasechnik:2009bq,Coughlin:2009tr,HarlandLang:2010ep,HarlandLang:2010ys,HarlandLang:2010te,HarlandLang:2011qd,Lebiedowicz:2011nb,Maciula:2011iv,Machado:2011vh} and experimentally~\cite{Albrow:2010zz,Abulencia:2006nb,cdf:2007na,Aaltonen:2009kg,Aaltonen:2009cj,Aaltonen:2011hi,Aaltonen:2007hs,Abazov:2010bk,Hubacek:2010zzb,Guryn:2008ky,Lee:2010zzp,Guryn:2011zz,Schicker:2011ug,LHCb,LHCb1,LHCb2,Royon:2010sm,Staszewski:2011bg} (see~\cite{Albrow:2009nj,Royon:2008ff} for reviews). The CEP of an object $X$ may be written in the form
\begin{equation}\label{cep}
pp(\overline{p}) \to p+X+p(\overline{p})\;,
\end{equation}
where $+$ signs are used to denote the presence of large rapidity gaps. An important advantage of these reactions is
 that they provide an especially clean environment in which to measure the nature and quantum numbers (in particular, the spin and parity) of new resonance states, from `old' Standard Model mesons to Beyond the Standard Model Higgs bosons (see for instance \cite{Kaidalov:2003fw,Kaidalov:2003ys,HarlandLang:2010ep,Heinemeyer:2007tu,Heinemeyer:2010gs,Khoze:2010ba,Chaichian:2009ts} and references therein). One of the most interesting examples is the CEP of the Higgs boson, which is at the heart of the FP420 LHC upgrade project~\cite{Albrow:2008pn}:   through the installation of dedicated forward proton detectors 420m away from the ATLAS and/or CMS detectors, it is hoped that detailed studies of new physics in high--luminosity runs of the LHC can be performed.

As discussed in detail in~\cite{HarlandLang:2010ep,Khoze:2004yb,Khoze:2004ak}, the CEP of, for instance, $\gamma\gamma$, heavy $(c,b)$ quarkonia, new charmonium--like states or meson pairs with sufficiently large $p_\perp$ can serve as `standard candle' processes with which we can benchmark predictions for new CEP physics at the LHC, as well as offering a promising way to study various aspects of QCD. In particular, studies of exclusive meson pair production provide a novel test of the perturbative CEP formalism (see e.g.\cite{Khoze:2001xm,Khoze:2008cx}), with all its non--trivial ingredients, from the structure of the hard subprocess to the incorporation of various screening effects caused by absorption, see~\cite{Dokshitzer:1991he,Bjorken:1992er}.

It is expected that a large CEP data sample will be available in the near future from measurements performed by STAR at RHIC~\cite{Guryn:2008ky,Lee:2010zzp,Guryn:2011zz}, where roman pot (RP) proton detectors are ideally positioned for observing the CEP of heavy quarkonia, charmonium--like resonances and charged meson pairs\footnote{In the special low--pile--up LHC runs, the TOTEM~\cite{Anelli:2008zza} and ALFA \cite{:2008zzj} detectors could study the CEP of hadronic systems (induced by the fusion of two soft Pomerons) using their RP proton taggers, placed at 220m and 240m from the LHC interaction points IP5 and IP1, respectively, in combination with the central detectors of CMS~\cite{cms-totem} and ATLAS~\cite{Staszewski:2011sk}. The expected cross sections (10--50 $\mu$b~\cite{Ryskin:2009tj,Ryskin:2011qh}) should be large enough to allow sufficient event rates during even short runs with special LHC optics. Of particular interest here is the possibility of measuring azimuthal correlations between the momenta of the outgoing protons, which presents a way to probe the proton opacity~\cite{Khoze:2002nf}. Recently TOTEM has reported~\cite{ken} encouraging preliminary results on central diffractive production measured with RP taggers. The possibility of measuring exclusive low mass pion pair production using the ALFA RP detectors during special low--luminosity LHC runs is also currently under discussion~\cite{Staszewski:2011bg}.}.

As noted in~\cite{HarlandLang:2011qd}, even without forward proton spectrometers, central diffractive processes of the type
\begin{equation}\label{dd}
pp(\overline{p}) \to Y+X+Z\;,
\end{equation}
with large rapidity gaps separating the centrally produced system $X$ from the products, $Y$ and $Z$, of the proton (antiproton) dissociation, are still of considerable practical and theoretical interest. If the incoming protons
are dissociated into low masses ($M_{Y,Z}\lesim$ 2 GeV), then these reactions continue to exhibit many of the attractive properties of CEP. These  central diffractive processes can be measured using forward rapidity gap triggers, with the help of simple scintillation (forward shower) counters (FSCs)~\cite{Albrow:2008az}. Such a strategy was successfully implemented at the Tevatron by CDF~\cite{Albrow:2010zz}, \cite{Abulencia:2006nb,cdf:2007na,Aaltonen:2009kg,Aaltonen:2009cj,Aaltonen:2011hi},  where a rapidity gap trigger was used to veto on particles with pseudorapidity $|\eta| <$ 7.4 on each side of the central system. The addition of FSCs to the LHC experiments could provide sufficient rapidity coverage, extending to the `blind' region $6<|\eta|
\lesssim 8$, which would allow the exclusion of events with high--mass and a large fraction of events with low--mass diffractive dissociation in the special low pile--up LHC runs. The FSC counters proposed for CMS~\cite{Albrow:2008az,cms} have recently been installed in the LHC tunnel and can be operational for low pile--up conditions which require special LHC running modes. A proposed dedicated CMS run with an integrated luminosity of about $100{\rm pb}^{-1}$ of single no--pile--up interactions, with FSC (and Zero Degree Calorimeters (ZDC) which detect neutral particles) engaged and taking data in common with TOTEM is especially promising\footnote{We thank Mike Albrow for a valuable discussion about this.}. It is also worth mentioning that by employing FSCs a rich CEP physics programme could also be realised with the LHCb~\cite{LHCb,LHCb1,LHCb2} experiment. The excellent particle identification and high momentum resolution of the LHCb detector are especially beneficial for measurements of low multiplicity final states. In the near future, LHCb is planning to perform measurements of exclusive meson pair production using the low multiplicity dihadron trigger, including $\chi_c$ decays into $\pi^+\pi^-$ and $K^+K^-$ states~\cite{moran2,LHCb}. Searches at LHCb for the baryonic ($p\overline{p}$, $\Lambda\overline{\Lambda}$)
decay channels of the $\chi_c$ mesons and new charmonium--like states are also now under discussion~\cite{Barsuk:2012ic}. A promising study of low central mass CEP events is also ongoing at ALICE~\cite{Schicker:2011ug}, using additional scintillator detectors placed on both sides of the central barrel, which allows tagging of double rapidity gap events.

In 2007 CDF published a search for $\gamma\gamma$ CEP~\cite{cdf:2007na} at the Tevatron, with $E_\perp(\gamma) >$ 5~GeV. Three candidate events were observed, in agreement with the expectation of~\cite{Khoze:2004ak}. Subsequently, to increase statistics the $E_\perp(\gamma)$ threshold has been decreased to 2.5 GeV, and in~\cite{Aaltonen:2011hi} the  observation of 43  $\gamma\gamma$ events in $|\eta(\gamma)|<1.0$ with no other particles detected in $-7.4<\eta<7.4$ is reported, which corresponds to a cross section of $\sigma_{\gamma\gamma}= 2.48^{+0.40}_{-0.35} $ $({\rm stat})^{+0.40}_{-0.51}$ $ ({\rm syst})$ pb. The theoretical cross section, calculated using the formalism described in~\cite{HarlandLang:2010ep,Khoze:2004ak} and implemented in the SuperCHIC MC generator~\cite{SuperCHIC}, is 1.42 pb using MSTW08LO PDFs~\cite{Martin:2009iq} and 0.35 pb using MRST99 (NLO) PDFs~\cite{Martin:1999ww}, while the $p_\perp$, $\Delta \phi$ and invariant mass distributions of the $\gamma\gamma$ pair are well described by the MC. In the analysis in~\cite{Aaltonen:2011hi} special attention was paid to the possible background caused by $\pi^0\pi^0$ CEP, since one or both of the photons from $\pi^0 \to \gamma\gamma$ decay can mimic the `prompt' photons from $gg \to \gamma\gamma$ CEP. Importantly, CDF has found that the contamination caused by $\pi^0\pi^0$   CEP is very small ($< 15$ events, corresponding to a ratio $N(\pi^0\pi^0)/N(\gamma\gamma)<0.35$, at 95\% CL). This supports the non--trivial theoretical result of~\cite{HarlandLang:2011qd}, calculated within a perturbative QCD framework, which predicts $\sigma(\pi^0\pi^0)/\sigma(\gamma\gamma)\approx 0.01$ for the CDF event selection\footnote{We recall that, due to the $J_z=0$ `selection rule' which operates for CEP, the LO $gg \to \pi^0\pi^0$ amplitude vanishes for $J_z=0$ incoming gluons, leading to a strong suppression in the production cross section.}. This adds another strong argument in favour of further developing a better quantitative theoretical understanding of the CEP of hadron pairs ($h=\pi,K,p,\Lambda$)
\begin{equation}\label{hh}
pp(\overline{p}) \to p+h\overline{h}+p(\overline{p})\;
\end{equation}
in the spirit of the discussion in~\cite{HarlandLang:2011qd}. We recall that these channels, especially $\pi^+\pi^-$ and $K^+K^-$, are also ideally suited for the spin--parity analysis of the $\chi_c$ mesons, and detailed studies of the CEP of such states provide a promising way to probe various aspects of QCD, see for example~\cite{HarlandLang:2010ep,HarlandLang:2010ys,HarlandLang:2010te,HarlandLang:2011qd}, \cite{Khoze:2004yb}.

Recently LHCb has reported preliminary results on the CEP of $\chi_{c}$ mesons in the $\chi_c\to J/\psi\,+\,\gamma$ channel, where vetoing was imposed on additional activity in the rapidity region $1.9<\eta<4.9$, with sensitivity to charged particles in the backwards region $-4<\eta<-1.5$~\cite{LHCb,LHCb1,LHCb2,moran2}. While the $\chi_{c(0,1)}$ production data are in good agreement with the theoretical predictions for exclusive
production~\cite{HarlandLang:2009qe,HarlandLang:2010ep}, the observed $\chi_{c2}$ rate is somewhat higher. However, it is worth recalling that the observed LHCb data include some fraction of events with proton dissociation, as in (\ref{dd}). In~\cite{HarlandLang:2011qd} qualitative arguments were given that the protons dissociative process should favour the production of higher spin $\chi_{c(1,2)}$ states, with the $\chi_{c2}$ yield being particularly enhanced. Recall that in the non--exclusive case the momentum $k_\perp$ transferred through the $t$--channel $gg$ system (see Fig.~\ref{fig:pCp}) can be rather large, which leads to an increase of the higher spin $\chi_{c(1,2)}$ contributions. The $\chi_{c2}$ central production cross section, which is proportional to $k_\perp^4$, is in particular expected to be sizeably enhanced. However a more accurate account of the effects caused by the un--instrumented regions in the LHCb experiment \cite{LHCb,LHCb1,LHCb2} requires more detailed quantitative studies\footnote{On the experimental side, the addition of FSCs on both sides of the LHCb experiment \cite {Lamsa:2009ej} would allow a more efficient veto on inelastic events and should greatly clarify the situation.}.

We note that CMS has published~\cite{Chatrchyan:2011ci} results of a measurement at 7 TeV during the 2010 LHC run of exclusive dimuon production mediated by two--photon fusion. This could be an important step towards forthcoming CEP studies with the CMS detector at the LHC. The observation of $\gamma\gamma$ CEP events, a search for which was recently reported by CMS in~\cite{CMSgam}, would in particular be of much interest\footnote{We recall that such a measurement, and indeed measurements of other CEP processes more generally, will however become increasingly difficult during the current high luminosity LHC runs, as the pile--up rate prevents an effective use of the rapidity gap veto method. As described above, this problem may in principle be overcome by making use of proton taggers, which may be viable in the case of for example $\pi\pi$ CEP~\cite{Staszewski:2011bg}, but in the case of $\gamma\gamma$ CEP this may prove very challenging due to the quite low expected cross sections.}.

The high precision of the new CDF data on $\gamma\gamma$ CEP~\cite{Aaltonen:2011hi} and the expectation of new results from the RHIC and LHC experiments on central diffractive production encourages interest in a more detailed theoretical understanding of these processes, and in reducing the theoretical uncertainties. Apart from the dependence on the choice of the set of parton distribution functions (PDFs) used for the computation of the CEP cross sections, the main uncertainties in the theoretical predictions come from possible NLO effects\footnote{At present the predictions in \cite{HarlandLang:2009qe},\cite{HarlandLang:2010ep,HarlandLang:2010ys,HarlandLang:2010te,HarlandLang:2011qd} are based mainly on the LO formulae.} and uncertainties in the evaluation of the soft absorptive effects. In this paper we revisit these theoretical uncertainties, paying special attention to the structure of the NLO corrections, in particular for the case of $\gamma\gamma$ CEP. Based on the most recent Durham models of soft diffraction~\cite{Ryskin:2009tj, Ryskin:2011qh}, we update the expectations for both the eikonal and semi--enhanced absorptive corrections. We also present numerical results for LHC and RHIC energies accounting for both the expected experimental cuts and acceptances.

The $\pi\pi$ CEP process, mediated by Pomeron--Pomeron fusion, has been a subject of theoretical studies within a Regge--pole framework since the 1970s (see, for instance \cite{Pumplin:1976dm,Azimov:1974fa,Desai:1978rh} for early references and \cite{Lebiedowicz:2011nb,HarlandLang:2010ys,HarlandLang:2010te} for more recent ones). There have also been a variety of experimental results on low mass meson pair CEP, in particular from the CERN--ISR, with the cross sections in broad agreement with the expectations from Regge phenomenology (see~\cite{Albrow:2010yb,Albrow:2010zz} for reviews). As discussed in~\cite{HarlandLang:2010ys,HarlandLang:2010te}, at comparatively large meson transverse momenta $k_\perp$ CEP should be dominated by the perturbative 2--gluon exchange mechanism, where the $gg\to \pi \pi,K^+K^-$ subprocess can be modelled using a generalisation of the
formalism of~\cite{Brodsky:1981rp,Chernyak:2006dk}. In the experimentally relevant kinematic region ($M_{\pi\pi}\sim M_\chi$, $k_\perp(\pi)\sim M_\chi/2$) in principle both (non--perturbative and perturbative) mechanisms could contribute to the overall rate and this issue requires careful investigation. Of special interest is the transition region between these two contributions which is very sensitive to the behaviour of the meson form factor $F_M(t)$.
Here we perform a detailed study of the CEP of meson pairs paying special attention to the perturbative regime, and the transition between the non--perturbative and perturbative regimes.

Finally, we note that for the sake of brevity we will not discuss the case of exclusive dijet production here (see for instance~\cite{Martin:1997kv,Khoze:2006iw}). Although it is certainly an interesting channel through which we can test the pQCD CEP framework out to quite high $M_{jj}$, it is not without issues. In particular, in the dijet case it is hard to define purely exclusive kinematics and to exclude the contribution of events with a relatively soft third jet~\cite{Khoze:2006iw}. There are also additional uncertainties to consider, in particular in the value of jet's $E_T$, caused by the details of the jet searching algorithm.

This paper is organized as follows. In Section 2 we recall the main aspects of the perturbative formalism for CEP. In Section~3 we consider in some detail the impact of NLO corrections and the choice of parton distribution functions on our CEP predictions, using $\gamma\gamma$ CEP as a paradigm `standard candle' process. In Section~4  we discuss the proton dissociation contribution to standard `elastic' CEP, which is relevant for when CEP processes are identified by large rapidity gaps, rather than by explicit tagging of the final--state protons. In Section~5 we expand on our previous treatment of meson--pair CEP by considering two other potentially important contributions, the first (perturbative) arising from the so--called `symmetric' contribution and the second (non--perturbative) from double Pomeron exchange. In Section~6 we present and discuss our numerical predictions for meson pair and $\gamma\gamma$ CEP at the LHC, Tevatron and RHIC, and in Section~7 we summarize our results and make some concluding remarks.

\section{Central exclusive production}\label{CEPform}

\begin{figure}[b]
\begin{center}
\includegraphics[scale=1.0]{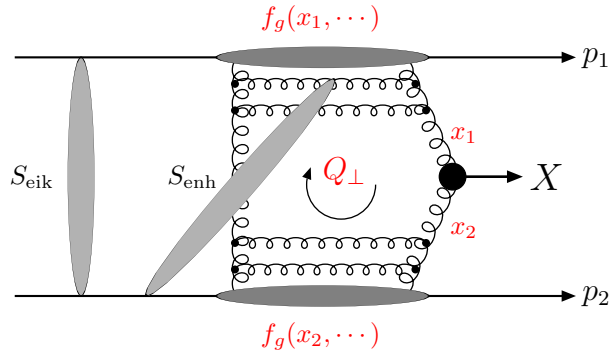}
\caption{The perturbative mechanism for the exclusive process $pp \to p\,+\, X \, +\, p$, with the eikonal and enhanced survival factors 
shown symbolically.}
\label{fig:pCp}
\end{center}
\end{figure} 

The formalism used to calculate the perturbative CEP cross section is explained in detail elsewhere~\cite{Khoze:2001xm,HarlandLang:2010ep,Khoze:2000jm,Khoze:2000cy} and so we only review the relevant aspects here. The amplitude is described by the diagram shown in Fig.~\ref{fig:pCp}, where the hard subprocess $gg \to X$ is initiated by gluon--gluon fusion and the second $t$--channel gluon is needed to screen the colour flow across the rapidity gap intervals. We can write the `bare' amplitude in the factorised form~\cite{Kaidalov:2003fw,Khoze:2001xm,Khoze:2004yb,Kaidalov:2003ys,Khoze:2004ak}
\begin{equation}\label{bt}
T=\pi^2 \int \frac{d^2 {\bf Q}_\perp\, \mathcal{M}}{{\bf Q}_\perp^2 ({\bf Q}_\perp-{\bf p}_{1_\perp})^2({\bf Q}_\perp+{\bf p}_{2_\perp})^2}\,f_g(x_1,x_1', Q_1^2,\mu^2;t_1)f_g(x_2,x_2',Q_2^2,\mu^2;t_2) \; ,
\end{equation}
where the $f_g$'s in (\ref{bt}) are the skewed unintegrated gluon densities of the proton: in the kinematic region relevant to CEP, they are given in terms of the conventional (integrated) densities $g(x,Q_i^2)$. $t_i$ is the 4--momentum transfer squared to proton $i$ and $\mu$ is the hard scale of the process, taken typically to be of the order of the mass of the produced state: as in~\cite{Khoze:2004yb,Kaidalov:2003ys}, we use $\mu=M_X/2$ in what follows. The $t$--dependence of the $f_g$'s is isolated in a proton form factor, which we take to have the phenomenological form $F_N(t)=\exp(bt/2)$, with $b=4 \,{\rm GeV}^{-2}$.  $\mathcal{M}$ is the colour--averaged, normalised sub--amplitude for the $gg \to X$ process
\begin{equation}\label{Vnorm}
\mathcal{M}\equiv \frac{2}{M_X^2}\frac{1}{N_C^2-1}\sum_{a,b}\delta^{ab}q_{1_\perp}^\mu q_{2_\perp}^\nu V_{\mu\nu}^{ab} \; .
\end{equation}
Here $a$ and $b$ are colour indices, $M_X$ is the central object mass, $V_{\mu\nu}^{ab}$ represents the $gg \to X$ vertex and $q_{i_\perp}$ are the transverse momenta of the incoming gluons, given by
\begin{equation}
q_{1_\perp}=Q_\perp-p_{1_\perp}\,, \qquad
q_{2_\perp}=-Q_\perp-p_{2_\perp}\,,
\label{qperpdef}
\end{equation}
where $Q_\perp$ is the momentum transferred round the gluon loop and $p_{i_\perp}$ are the transverse momenta of the outgoing protons. Only one transverse momentum scale is taken into account in (\ref{bt}) by the prescription
\begin{align}\nonumber
Q_1 &= {\rm min} \{Q_\perp,|({\bf Q_\perp}-{\bf p}_{1_\perp})|\}\;,\\ \label{minpres}
Q_2 &= {\rm min} \{Q_\perp,|({\bf Q_\perp}+{\bf p}_{2_\perp})|\} \; .
\end{align}
The longitudinal momentum fractions carried by the gluons satisfy
\begin{equation}\label{xcomp}
\bigg(x' \sim \frac{Q_\perp}{\sqrt{s}}\bigg)  \ll \bigg(x \sim \frac{M_X}{\sqrt{s}}\bigg) \; ,
\end{equation} 
where $x'$ is the momentum fraction of the second $t$-channel gluon. The differential cross section at $X$ rapidity $y_X$ is then given by
\begin{equation}\label{ampnew}
\frac{{\rm d}\sigma}{{\rm d} y_X}=\langle S^2_{\rm enh}\rangle\int{\rm d}^2\mathbf{p}_{1_\perp} {\rm d}^2\mathbf{p}_{2_\perp} \frac{|T(\mathbf{p}_{1_\perp},\mathbf{p}_{2_\perp}))|^2}{16^2 \pi^5} S_{\rm eik}^2(\mathbf{p}_{1_\perp},\mathbf{p}_{2_\perp})\; ,
\end{equation}
where $T$ is given by (\ref{bt}) and $S^2_{\rm eik}$ is the `eikonal' survival factor, calculated using a generalisation of the `two--channel eikonal' model for the elastic $pp$ amplitude (see ~\cite{Khoze:2002nf} and references therein for details).

Besides the effect of eikonal screening, $S_{\rm eik}$, there is an additional suppression caused by the rescatterings of the intermediate partons (inside the unintegrated gluon distribution, $f_g$). This effect is described by the so--called enhanced Reggeon diagrams and usually denoted as $S^2_{\rm enh}$, see Fig.~\ref{fig:pCp}. The value of $S^2_{\rm enh}$ depends mainly on the transverse momentum of the corresponding partons, that is on the argument $Q^2_i$ of $f_g(x,x',Q^2_i,\mu^2;t)$ in (\ref{bt}), and depends only weakly on the $p_\perp$ of the outgoing protons (which formally enters only at NLO). While in~\cite{HarlandLang:2010ep,HarlandLang:2009qe} the $S^2_{\rm enh}$ factor
was calculated using the formalism of~\cite{Ryskin:2009tk}, here, following~\cite{HarlandLang:2011qd}, we use a newer version of the multi--Pomeron model~\cite{Ryskin:2011qe} which incorporates the continuous dependence on $Q^2_i$ and not only three `Pomeron components' with different `mean' $Q_i$. We therefore in practice include the $S_{\rm enh}$ factor inside the integral (\ref{bt}), with $\langle S^2_{\rm enh}\rangle$ being its average value integrated over $Q_\perp$.

If we consider the exact limit of forward outgoing protons, $p_{i_\perp}=0$, then we find that after the $Q_\perp$ integration (\ref{Vnorm}) reduces to~\cite{Khoze:2001xm}
\begin{equation}\label{mprop}
\mathcal{M}\propto q_{1_\perp}^i q_{2_\perp}^j V_{ij} \to \frac{1}{2}Q_\perp^2(V_{++}+ V_{--})\sim\sum_{\lambda_1,\lambda_2}\delta^{\lambda_1\lambda_2}V_{\lambda_1\lambda_2}\;,
\end{equation}
where $\lambda_{(1,2)}$ are the gluon helicities in the $gg$ rest frame. The only contributing helicity amplitudes are therefore those for which the $gg$ system is in a $J_z=0$ state, where the $z$--axis is defined by the direction of motion of the gluons in the $gg$ rest frame, which, up to corrections of order $\sim q_\perp^2/M_X^2$, is aligned with the beam axis.  In general, the outgoing protons can pick up a small $p_\perp$, but large values are strongly suppressed by the proton form factor, and so the production of states with non--$J_z=0$ quantum numbers is correspondingly suppressed (see~\cite{HarlandLang:2010ep,HarlandLang:2009qe,HarlandLang:2011zz} for examples of this in the case of $\chi_{(c,b)}$ and $\eta_{(c,b)}$ CEP). In particular, we find roughly that
\begin{equation}\label{simjz2}
\frac{|T(|J_z|=2)|^2}{|T(J_z=0)|^2} \sim \frac{\langle p_\perp^2 \rangle^2}{\langle Q_\perp^2\rangle^2}\;,
\end{equation}
which is typically of order $\sim1/50-1/100$, depending on the central object mass, c.m.s. energy $\sqrt{s}$ and choice of PDF set. As discussed in~\cite{HarlandLang:2011qd}, this `$J_z=0$ selection rule'~\cite{Kaidalov:2003fw,Khoze:2000mw,Khoze:2000jm} will have important consequences for the case of meson pair CEP. Finally, we note that in (\ref{mprop}) the incoming gluon (transverse) polarizations are averaged over at the {\it amplitude} level: this result is in complete contrast to a standard inclusive production process where the {\it amplitude squared} is averaged over all gluon helicities. Eq. (\ref{mprop}) can be readily generalised to the case of non--$J_z=0$ gluons which occurs away from the forward proton limit, see in particular Section 4.1 (Eq. (41)) of~\cite{HarlandLang:2010ep}, which we make use of throughout to calculate the $M\overline{M}$ CEP amplitudes from the corresponding $gg\to M\overline{M}$ helicity amplitudes.

\section{$\gamma\gamma$ CEP and NLO corrections}\label{gamnlo}

\subsection{$\gamma\gamma$ CEP revisited}\label{gamrev}

As discussed in the Introduction, in~\cite{cdf:2007na} CDF published a search for $\gamma\gamma$ CEP with $E_\perp(\gamma) >$ 5 GeV, observing three candidate events. Subsequently, to increase statistics the $E_\perp(\gamma)$ threshold was decreased to 2.5 GeV, and in~\cite{Aaltonen:2011hi} the observation of 43 candidate $\gamma\gamma$ events in $|\eta(\gamma)|<1.0$ with no other particles detected in $-7.4<\eta<7.4$ was reported, which corresponds to a cross section of $\sigma_{\gamma\gamma}= 2.48^{+0.40}_{-0.35} ({\rm stat}) {}^{+0.40}_{-0.51} ({\rm syst})$ pb, and with a $\pi^0\pi^0$ contamination consistent with zero. The theoretical cross section, calculated using the formalism described in~\cite{Khoze:2004ak,HarlandLang:2010ep} and implemented in the SuperCHIC MC generator~\cite{SuperCHIC}, is 1.42 pb using MSTW08LO PDFs~\cite{Martin:2009iq} and 0.35 pb using MRST99 (NLO) PDFs~\cite{Martin:1999ww}. Evidently the prediction using the LO PDF set is consistent with the result within theoretical uncertainties, although both predictions lie below the observed cross section, particularly so in the case of the MRST99 PDF choice.

It is natural to ask why the $\gamma\gamma$ CEP cross section predictions in~\cite{HarlandLang:2010ep} are somewhat lower than the data. In fact, there are reasons why we may expect this to be the case. Most importantly\footnote{Experimentally we may also expect the observed $\gamma\gamma$ cross section to be enhanced by the small fraction of $\gamma\gamma$ events seen by CDF which are not truly exclusive, but rather due to double diffractive production where one or both of the proton and antiproton dissociates, but where the proton dissociation products are not seen within the CDF acceptance; we estimate such a fraction  to be 10\% or lower.}, we recall that the prediction of~\cite{HarlandLang:2010ep} includes only the LO perturbative QCD contribution to the $\gamma\gamma$ CEP process. In general, we may reasonably expect a numerically large NLO K--factor correction to the $gg \to X$ subprocess: for example, the higher order corrections to Standard Model Higgs boson production via $gg \to H$ (see~\cite{Spira:1995rr,Kunszt:1996yp} and references therein) and P--wave quarkonia decay $\chi \to gg$~\cite{Barbieri:1981xz} are known to be quite large\footnote{In the case of $\chi_{c,b}$ CEP we assume the same K--factor for the $\chi \to gg$ and $gg \to \chi$ cases, see~\cite{HarlandLang:2009qe,HarlandLang:2010ep}, which is only true to a certain degree of approximation.}. In the case of inclusive Higgs boson production via $gg$ fusion, the K--factor is approximately given by
\begin{equation}\label{higgsK}
 K \approx 1+\frac{\alpha_S(M_H^2)}{\pi} \bigg[\pi^2+\frac{11}{2}\bigg]\approx 1.5\;,
\end{equation}
for a light Higgs ($M_H\approx 125$ GeV). We can see in particular that the NLO cross section is numerically enhanced by a factor of $\pi^2$, the origin of which can be traced back to a Sudakov--like double logarithm, $\alpha_S \log^2(-\vert q^2 \vert /M_H^2)$, when the imaginary part of the logarithm is squared. Such an enhancement is also observed to occur in the NLO correction to the $\chi_{(c,b)0} \to gg$ decay. We may reasonably expect similarly enhanced NLO corrections to Higgs and $\chi_{(c,b)0}$ CEP, and in~\cite{Khoze:2001xm,Khoze:2004yb,HarlandLang:2009qe} such an assumption is made to estimate the relevant NLO cross sections, although we note that the K--factor in (\ref{higgsK}) is for the spin--averaged case, whereas in CEP only the $J_z=0$ helicity amplitudes are important. It is also worth mentioning that an additional positive correction (of the order of about $+20$\%) could come from a consistent account of the self--energy insertions in the propagator of the screening gluon, in the spirit of the discussion in~\cite{Shuvaev:2008yn}.

In the case of $\gamma\gamma$ CEP, the NLO corrections to the $g(\lambda_1)g(\lambda_2)\to \gamma(\lambda_3)\gamma(\lambda_4)$ subprocess helicity amplitudes are calculated in~\cite{Bern:2001df}. While it is quite difficult to trace the origin of all the $\pi^2$ terms in the reasonably complicated analytic expressions for the helicity amplitudes in~\cite{Bern:2001df}, our understanding is that the $\gamma\gamma$ NLO virtual corrections may be similarly enhanced, e.g. by Sudakov--like double logarithmic terms. Indeed, including the finite part of these NLO corrections~\cite{Bern:2001df} to the $gg\to\gamma\gamma$ subprocess, we find that the K--factor for the $\gamma\gamma$ CEP cross section is quite large, given by approximately $K\approx 1.6$ for experimentally relevant $M_{\gamma\gamma}$ values. However, as discussed in the following section, this does not constitute the {\it full} NLO K--factor, and therefore can only be taken as a guide. Nonetheless, this result is suggestive that the LO estimate for the $\gamma\gamma$ CEP cross section may receive sizeable positive higher order corrections.

This effect could easily give a factor of $\sim 2$ increase in the theoretical $\gamma\gamma$ CEP cross section, when compared to the calculation in~\cite{HarlandLang:2010ep}, thus alleviating some of the tension that exists between the theory and the latest CDF measurement. Nonetheless, in the case of the MRST99 PDF choice, the theoretical prediction is almost an order of magnitude below the experimental value, and so it appears that, even taking into account the other sizeable theoretical uncertainties as well as the issues discussed above, the $\gamma\gamma$ CEP cross section calculated with this PDF choice is too small. This may indicate that in fact the `true' gluon PDF in the low--$x$, low--$Q^2$ region relevant to CEP is better described by the LO MSTW08 fit, which gives a higher density of gluons due to the steeper $x$ dependence than the relatively flat MRST99 PDF. Of course, given the existing theoretical uncertainties in the CEP calculation, it is difficult to make a very strong statement about this, and certainly other measurements, in particular of $\gamma\gamma$ (and other processes) at the LHC, would be very useful in shedding light on this issue. 

We note that the proton's PDFs are physical quantities which are defined as the expectation values of the corresponding QCD operators over the proton state. There are no indices, such as LO, NLO, $\overline{\rm MS}$,... , in their definition. However, these PDFs are not measured directly, rather they are determined by optimising the description of some hard cross section by the convolution of the PDF with the corresponding coefficient function (i.e. the hard matrix element $|\mathcal{M}|^2$). At this stage the resulting PDF starts to depend on the perturbative approximation (LO, NLO,...) and the choice of factorisation and renormalisation schemes used for the calculation of $|\mathcal{M}|^2$. For example, for the Drell--Yan process the K--factor $K=|\mathcal{M}^{\rm NLO}|^2/|\mathcal{M}^{\rm LO}|^2 \sim 1.5 - 2$ is rather large, which will lead to a larger PDF using the LO $|\mathcal{M}|$ than when using the NLO $|\mathcal{M}|$. However we cannot claim  that different processes have the same K--factors. For example, in $\gamma\gamma$ CEP the NLO correction is expected to be numerically of the same order (for the hard $|\mathcal{M}|^2$) as that in the DY case but in CEP the value of $|\mathcal{M}|^2$ is multiplied by $(xg)^4$. 

In practice, global fits of the gluon density, $g(x,Q^2)$, at low $x$ are based almost entirely on measurements of $F_2 (x,Q^2)$, in particular through the effect that $g(x,Q^2)$ has on the DGLAP evolution of the quark PDF and therefore on ${\rm d}F_2(x,Q^2)/{\rm d}\ln Q^2$. The extracted gluon density therefore depends on the approximations used to calculate this evolution, but at sufficiently low scales the higher order and power (including the absorptive) corrections to the pure DGLAP evolution are certainly not negligible. In most global PDF analyses (performed within the linear, pure twist 2  perturbative DGLAP approach) cuts are imposed on the data (for example in $Q^2$ and/or $W^2$) in order to minimise the effect of these corrections, but if there is some residual contamination then this will simply be absorbed into the input gluon distribution. In addition, the input $g(x,Q_0^2)$ is adjusted at low $x$ in order to obtain a satisfactory fit to data over a wide $Q^2$ and $x$ range, over most of which the higher--twist and higher--order corrections are expected to be small. Although this approach may therefore provide a satisfactory description of the gluon PDF at large $Q^2$, it may give unreliable results for very low $Q^2$ and $x$. Indeed in the case of modern NLO $\overline{\rm MS}$ PDF sets, the fits tend to prefer a negative gluon, for example $g(x,Q_0^2 = 1\, {\rm GeV}^2) < 0$ for $ x \lesim 10^{-2}$ in the case of MSTW08 NLO. On the other hand there are some indications, for instance from $J/\psi$ diffractive photoproduction data at HERA (which is not currently used in global PDF analyses), that the `true' $g(x,Q^2)$ at low $x$ and $Q^2\sim 2-3 $ ${\rm GeV}^2$ is larger than the current NLO PDFs~\cite{Martin:2007sb}.

In contrast, the LO PDFs, which generally have a $g(x,Q_0^2)$ which is positive at small $x$, do not give a particularly good  description of the HERA $F_2(x,Q^2)$ data, and may in general be too large at low $x$ and $Q^2$ in order to compensate for the non--negligible higher--order terms which are absent in the LO DGLAP evolution of the quark PDF.\footnote{In particular, at LO there is no $1/z$ singularity in the $P_{qg}(z)$ and  $P_{qq}(z)$ splitting function, which is present at NLO and higher orders.} 

It is therefore not unreasonable to expect that the `true' gluon PDF at low $x$ and $Q^2$, relevant to the relatively low--mass CEP processes that we are interested in here, lies somewhere between the lower and upper bounds set by the NLO and LO PDF sets, respectively. For this reason, we use the MSTW08 LO and MRST99 NLO sets in our study, arguing that they span a realistic range of small $x$ gluon distributions (we show this explicitly in Section~\ref{pdfcomp}). Of the available NLO PDFs, we have chosen to use the older MRST99 set as our benchmark as this has a more benign small--$x$ form (in particular, it is somewhat flat at low $x$, $Q^2$), and will therefore not produce numerical instabilities that could occur with more modern NLO fits, in which the gluon can be negative in this region (at least in the $\overline{\rm MS}$ factorisation scheme generally used in the fits). 

\subsection{Remarks on NLO corrections to CEP processes}

\begin{figure}[h]
\begin{center}
\includegraphics[scale=0.5]{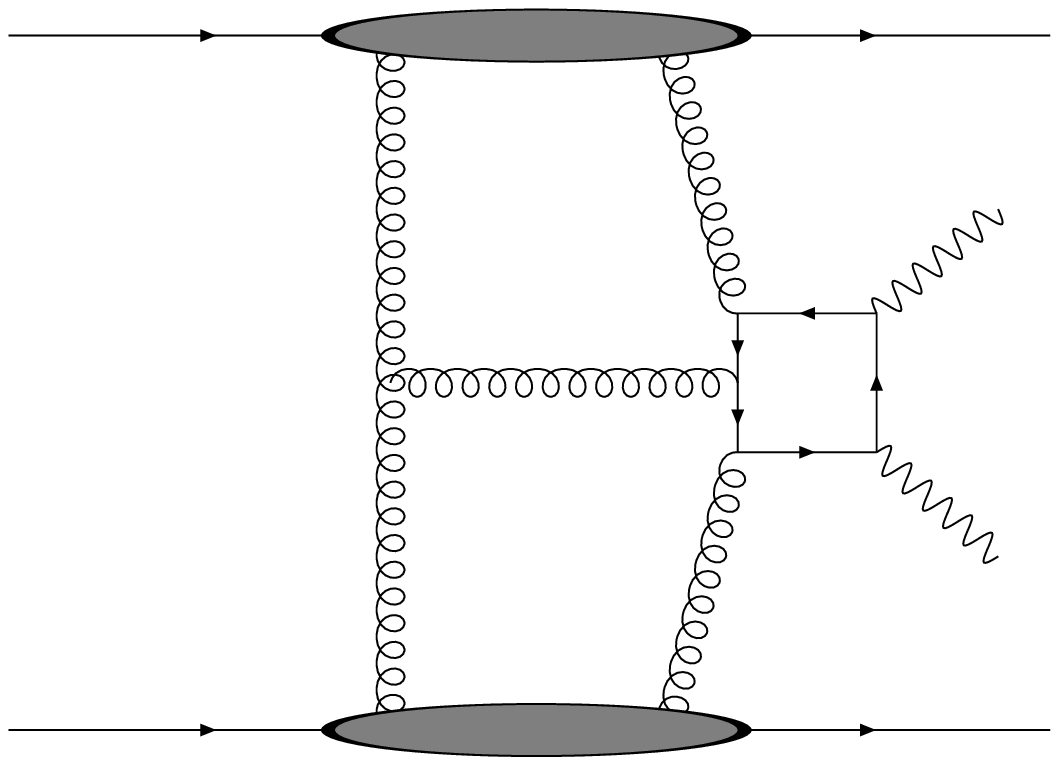}\qquad
\includegraphics[scale=0.5]{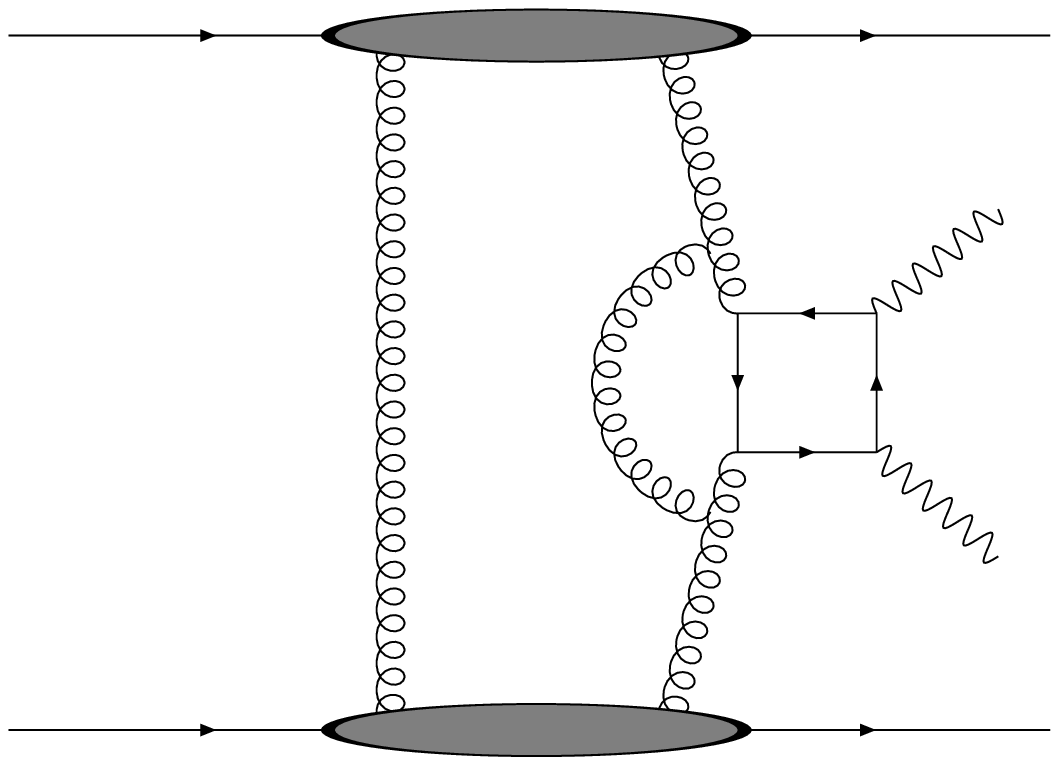}
\caption{Representative NLO virtual correction diagrams to the $\gamma\gamma$ CEP process.}\label{nlog}
\end{center}
\end{figure}

\noindent Considering the inclusive Higgs production case as a particular example, the K--factor given in (\ref{higgsK}), due to the finite part of the virtual corrections to the $gg \to H$ process, does not give the full NLO correction. For the fully inclusive Higgs production cross section we must also include the finite contribution from QCD radiation $gg \to Hg$ (as well as $q\overline{q} \to Hg$ and $qg \to Hq$) to get the correct total NLO correction. The net effect of all these contributions can in principle be quite different from the value given in (\ref{higgsK}), although in fact the fully inclusive cross section does receive a K--factor of a similar (large) size to this.

In the exclusive case the set of NLO diagrams that we must calculate is different. There are still the standard virtual NLO corrections to the `hard' $gg\to X$ matrix element ($X=H, \gamma\gamma...$), which can be calculated in the usual way or  taken from known NLO results, such as (\ref{higgsK}) in the case of Higgs production. However, in CEP the contributions due to real emission (which cannot occur) are no longer present, and thus there is no explicit cancellation of the infrared divergences present in the virtual loop amplitudes. Here, these are cancelled by a corresponding divergence hidden in the unintegrated parton distributions, $f_g$ (see (\ref{bt})). In particular, such a cancellation determines the lower $k_\perp$ limit in the Sudakov factor integral,
\begin{equation}\label{ts}
T_g(Q_\perp^2,\mu^2)={\rm exp} \bigg(-\int_{Q_\perp^2}^{\mu^2} \frac{d {\bf k}_\perp^2}{{\bf k}_\perp^2}\frac{\alpha_s(k_\perp^2)}{2\pi} \int_{0}^{1-\Delta} \bigg[ z P_{gg}(z) + \sum_{q} P_{qg}(z) \bigg]dz \bigg) \; .
\end{equation}
This occurs because some of the NLO loop diagrams in the hard matrix element are identical to those contributing to the Sudakov factor, included in the $f_g$'s. In particular, in the Feynman gauge one of the first, order $\alpha_s$, diagrams contributing to the Sudakov factor is due to gluon exchange between the two incoming partons in the hard subprocess, shown in Fig.~\ref{nlog} (right) for the case of $\gamma\gamma$ CEP. As the NLO matrix element contains exactly the same diagram, to avoid double counting we must exclude this contribution from the NLO matrix element. This is just the infrared divergent contribution caused by the `absence' of soft gluon emission, and thus in the NLO matrix element we should only include the region where the scale is larger than the argument $Q$ of $f_g(Q,\cdots)$, while the `infrared sensitive' low--$k_\perp$ part is included in the unintegrated parton densities, where there is a cancellation between diagrams in which the low $k_\perp$ gluon is absorbed by the active (right) $t$--channel gluon and the soft (left) screening  gluon in Fig.~\ref{fig:pCp}; the long--wavelength low $k_\perp<Q_\perp$ gluon is only sensitive to the whole colour charge of the $t$--channel {\em colourless} two--gluon state. As  was shown in \cite{Khoze:2008cx}, for large subenergies squared $\hat s=M^2_X\gg Q^2_\perp$, i.e. in the BFKL limit, such a cancellation may be accounted for by taking the lower limit $k_\perp>Q_\perp$ in the $k_\perp$ integral for the Sudakov form factor.

In the exclusive case, we must explicitly calculate these additional diagrams, the divergent parts of which are included in the $f_g$'s (in particular, the `Sudakov' loop diagram described above, and diagrams where additional gluons couple to the quark lines inside the hard matrix element, see for example Fig.~\ref{nlog} (left)), as well as those in which the new (additional) gluon couples to the `soft' screening gluon. These will all give finite contributions to the CEP NLO K--factor which will in general be different in the Higgs case from those contributing to (\ref{higgsK}), and these will also depend on the object, X, being produced. While the single and double logarithmically enhanced contributions from the higher order corrections to the CEP amplitude are already included in the Sudakov factor, the finite contributions coming from a full NLO calculation are not, and such a calculation has not yet been performed (although a subset of the contributing diagrams, in particular those which are logarithmically enhanced and therefore give a non--zero contribution to the Sudakov factor, were calculated in the Higgs case in~\cite{Coughlin:2009tr}). 

Finally, a few more comments about the NLO corrections to the CEP process are in order. First, we note that in an NLO calculation the unintegrated parton densities, $f_g$, used in (\ref{bt}) should be calculated to NLO accuracy (as described in \cite{Martin:2009ii}). Secondly, as mentioned above, there are contributions to the NLO cross section from diagrams in which the additional gluon is emitted from the `soft' screening gluon but absorbed somewhere in the `hard' matrix element: a representative diagram in the case of $\gamma\gamma$ CEP is shown in Fig.~\ref{nlog} (left). However, in general, when the hard scale $\mu\sim M_X$ is large, such a contribution should be treated as a power correction, which can be seen as follows (see also the discussion in~\cite{Khoze:2009er}). We will consider two extreme cases: when the additional gluon emitted from the screening (left) gluon is rather soft and has transverse momentum $q_\perp\sim Q_\perp\ll M_X$ and when the gluon has a large $q_\perp\sim \mu$. In the former case, the phase volume in the $\int {\rm d}q^2_\perp$ integral ($\sim Q^2_\perp$) cannot compensate the large virtuality of the additional propagator in the `hard' matrix element, which absorbs the gluon, and the expected contribution will be of the order of $Q^2_\perp/\mu^2$. In the latter case, the new gluon has a large $q_\perp\sim \mu$, which would mean that the scale and the transverse momenta of the unintegrated partons is also large, with the value of $q_\perp\sim \mu$ compensated by some other parton in $f_g$. On the other hand the integral (\ref{bt}) over $Q_\perp$ is superconvergent, i.e. it has the form
\begin{equation}\label{sp}
\int d^2Q_\perp
\frac{f_g(x_1,Q_\perp,\mu,...)f_g(x_2,Q_\perp,\mu,...)}{Q^4_\perp}\; .
\end{equation}
Normally this has a saddle point at $Q_\perp=Q_{\rm sp}\ll \mu$. Thus the contribution from the $Q_\perp\sim \mu$ domain will be suppressed by the power factor $Q^2_{\rm sp}/\mu^2$.

There is also some correction caused by the fact that in the `hard' matrix element the incoming gluons are off--mass--shell. This also leads to a power suppressed correction of order $Q^2_\perp/\mu^2$ since the virtuality of the incoming gluon is given by $Q^2_\perp\sim Q^2_{sp}\ll\mu^2$. However there is some contribution from the region $Q_\perp\sim \mu$ which, strictly speaking, can not be considered separately from the NLO calculation. This kinematical effect should be considered together with all other NLO diagrams, especially the diagrams which describe the creation of a highly off--mass--shell gluon from a previously low--virtuality parton, which may be treated as on--mass--shell. We should recall in particular that the matrix element with off--shell gluons is not gauge invariant and may contain some additional non--physical singularities which are cancelled in the sum of the diagrams which forms the gauge invariant group (or subgroup) of diagrams. 

In processes with large $M_X\gg Q_{\rm sp}$, such as Higgs boson CEP, these power corrections are numerically small and may be neglected. However at relatively low scales, as is the case for $\chi_c$ production or $\gamma\gamma$ production when $E_\perp(\gamma)$ is not too large (as in the CDF event selection), these corrections may give a non--negligible contribution, although without performing an explicit NLO calculation it is difficult to quantify this statement. We will return to this issue, in particular that of the initial gluon off--shellness, in the case of meson pair CEP in Section~\ref{symc}.

\section{On the role of proton dissociation}\label{dcep}
In current LHC measurements of central production only some restricted large rapidity gap (LRG) conditions can be imposed, leaving certain regions in the forward (backward) directions uninstrumented. As discussed in the Introduction (see also \cite{HarlandLang:2011qd}), without forward proton spectrometers there is always a possibility that an `exclusive' event does not come from the purely {\it elastic} process (\ref{cep}) but rather from central diffractive production (\ref{dd}), with sufficiently large rapidity gaps separating the  system $X$ from the products, $Y$ and $Z$, of the proton dissociation\footnote{Note that as shown in~\cite{Albrow:2008az,cms}, in the low pile-up runs ZDC detectors would allow a reduction in the contribution from  diffraction dissociation by at least a factor of 2.}. We shall use the nomenclature Pseudo--CEP (PCEP) to denote such central production with the dissociation of (one or two) incoming protons.

To evaluate the possible admixture due to proton dissociation we can use a formalism similar to that described in Section~\ref{CEPform} for the CEP case, where the outgoing protons remain intact. To achieve this we must replace the unintegrated gluon density $f_g(x_i,x'_i,Q^2_i,\mu^2;t)$ in (\ref{bt})  by another function $f_g(x_i,x'_i,Q^2_i,\mu^2;t;M^2_Y)$, which accounts for the proton dissociation into a state with mass $M_Y$. The problem is that the form of this function is practically unknown, and so the best we can do is to make some plausible assumptions about its behaviour.
 
For the case of low mass dissociation ($p\to N^*$), it is reasonable to assume the same $x,\, Q^2,\,\mu$ and $t$ behaviour as for the `elastic', $p\to p$,  process. That is, to incorporate low mass dissociation it is sufficient to multiply the original (purely exclusive) result by some factor $1+c$, where $c$ is the probability of proton dissociation into the relatively low mass state. These probabilities were measured at lower (fixed target and CERN--ISR) energies (see~\cite{Kaydalov:1971ta}, and ~\cite{Kaidalov:1979jz,Goulianos:1982vk} for reviews), and can be extrapolated to LHC energies by accounting for stronger absorptive effects, using the models for soft diffraction discussed in~\cite{Ryskin:2011qe,Ryskin:2009tk}. This gives $c\simeq 0.15$ for the overall probability of proton dissociation to states with mass $M_Y<2 $ GeV, in agreement with the analysis of~\cite{Kaydalov:1971ta}. A similar value of $c\simeq 0.2$ was found at HERA, by comparing the size of the diffractive DIS cross section measured using the leading proton spectrometer with that found by requiring a LRG in the forward direction\cite{hera,Chekanov:2008fh,Aaron:2012ad}. Thus to account for the possibility of low mass dissociation of both protons we have to increase the CEP prediction by a factor $(1+c)^2\sim 1.4$, where we take $c=0.2$ for definiteness.

The situation with high mass dissociation is more complicated. Usually, dissociation into high mass states is described in terms of triple--Pomeron diagrams. This contribution is driven by the triple--Pomeron vertex, $g_{3P}$ (see for example~\cite{Kaidalov:1979jz} for a review). For fixed momentum transferred through the Pomeron, $t=-p^2_\perp$, that is through the $t$--channel state formed by the gluons $x$ and $x'$ in Fig.~\ref{fig:pCp}, the ratio of
the high mass dissociation to the `elastic' contribution can be written as
\begin{equation}\label{el-dif}
R=\sigma(p\to M_Y)/\sigma(CEP)=\int\frac{dM^2_Y}{M^2_Y}\frac{g_N(0)g_{3P}(t)}{\pi g^2_N(t)} \; ,
\end{equation}
where the integral runs over the $M_Y$ interval available in a given experiment. The value of the proton--Pomeron vertex, $g_N(t)$, is known from elastic $pp$ scattering, while the triple--Pomeron vertex, $g_{3P}$, can be extracted from the description of lower energy  (CERN--ISR, Tevatron) data in the triple--Regge region, see for example~\cite{Kaidalov:1973tc}. After accounting for the eikonal--like absorptive corrections, the recent fit in~\cite{Luna:2008pp} gives $g_{3P}(0)=0.2g_N(0)$. However here we face a problem, as we will now discuss. Absorptive effects strongly depend on the shape of the triple--Pomeron amplitude in impact parameter, $b_t$, space. At lower $b_t$ the proton optical density is large, and absorption is much stronger. On the other hand, we only poorly know the $t$--slope, $b_{3P}$ (that is the $b_t$ size), of the $g_{3P}\propto \exp(b_{3P}t)$ vertex. Fits to the lower energy triple--Regge data~\cite{Kaidalov:1973tc,Luna:2008pp} indicate that the slope $b_{3P}$ is rather small, and may even be consistent with zero, with $b_{3P}<2$ GeV$^{-2}$.  Thus the value of the corresponding survival factor $S^2$ is uncertain. Moreover, when  we do not detect the outgoing proton (or the $M_Y$ state) we have to integrate over the squared momentum transfer $t$. In the `elastic' CEP case the integral is limited by the proton form factor, and the average $p_\perp$ is small. In `soft' single dissociation (corresponding to the triple--Pomeron diagram without any `hard' subprocess), the squared momentum $t$ goes through the LRG and is again limited by the proton form factor. This is not the case for proton dissociation in PCEP where the momentum $p_\perp$ goes across the loop between the `hard' subprocess matrix element $|\mathcal{M}|^2$, which may be viewed as a `point--like' blob of a very small size in $b_t$ space, and the $g_{3P}$ vertex. Now, according to (\ref{el-dif}), in the case of high mass dissociation we can arrive at very large values of $p^2_\perp\sim 1/b_{3P}$. This will lead to an unacceptably large probability for dissociation, as for large $p_\perp$ we cannot justify the factorization $f_g(x,...\mu^2;t,M^2_Y)=G(t)f_g(x,...\mu^2;M^2_Y)$ of the $t$ dependencies. Simultaneously, large values of $p_\perp$ will allow an increasing violation of the $J_z=0$ selection rule which operates for pure CEP; the admixture of the $|J_z|=2$ contribution (\ref{simjz2}) will increase like $\sim \langle p_\perp^2 \rangle^2$. This may be crucially important for $\chi_{c2}$ or $\pi\pi$ production where the $J_z=0$ CEP amplitudes are strongly suppressed (the $J_z=0$ amplitude is zero for $\chi_{c2}$ CEP in the non--relativistic quarkonium approximation, and zero at LO for $\pi\pi$ CEP)\footnote{As was mentioned in \cite{HarlandLang:2011qd}, an admixture due to a proton dissociation contribution could explain, at least in part, why the observed LHCb \cite{LHCb} $\chi_{c2}$ PCEP cross section is higher than the CEP prediction, although cuts on the central $\mu^+\mu^-$ system $p_\perp <0.9$ GeV are imposed to suppress this contribution. With the increased statistics that should hopefully be available from future data, a closer study of the dependence of the relative $\chi_c$ cross sections on this $p_\perp$ cut would be possible, shedding further light on this issue.}. 

Taking $b_{3P}=1 \,{\rm GeV}^{-2}$ we then integrate (\ref{el-dif}) up to the maximum $M_Y$ allowed by the uninstrumented $\Delta y$ for the relevant detector, and we evaluate the admixture of high mass dissociation in PCEP events selected by a rapidity gap veto to be about $C\simeq 30 - 40$\% for the CMS/ATLAS experiment and up to $ C\simeq 50$\% in the LHCb case\footnote{In the limit that the `hard' subprocess has characteristic  $E_\perp \gg 1$ GeV ($b_{3P}*E_\perp^2\gg 1$) this estimate is process independent. However, for e.g. $\chi_{c0}$ production the $gg \to \chi_{c0}$ vertex may effectively enlarge the total $t$-slope in the loop somewhat, leading to a smaller probability of dissociation.}. In other words, the enhancement factor by which we have to multiple the CEP cross section in order to estimate the PCEP rate is $\simeq(1+c +C/2)^2$, where the factor of `$1/2$' accounts for the fact that $C$ is the probability that either proton dissociates. We should recall, however, that there is a large uncertainty in this estimate, as discussed above\footnote{Further modification of the admixture of dissociation will arise after accounting for detector effects.}, and that this high mass dissociation will generate a rather large transverse momenta $p_\perp$ of the central system and, therefore, a larger admixture of the $|J_z|=2$ contribution.

\begin{table}
\begin{center}
\begin{tabular}{|l|c|c|c|}
\cline{1-3}
Single & Low $M_{Y,Z}$ ($\lesssim 2.5$ GeV)&High $M_{Y,Z}$ ($\gtrsim 2.5$ GeV)&\multicolumn{1}{c}{}\\
\cline{1-3}
$S^2$&$0.86 \pm 0.03$&$0.81 \pm 0.03$&\multicolumn{1}{c}{}\\
\hline
Double& (Low $M_Y$, Low $M_Z$) & (Low $M_Y$, High $M_Z$)&(High $M_Y$, High $M_Z$)\\
\hline
$S^2$&0.3 -- 0.45&0.2 -- 0.28&0.08 -- 0.16\\
\hline
\end{tabular}
\caption{Estimates for the survival factor (due to strong interactions between the colliding protons) for single ($pp\to p(Y)\,+\,\gamma\gamma\,+\,Z(p)$) and double ($pp\to Y\,+\,\gamma\gamma\,+\,Z$) diffractive two--photon production of a lepton pair, at $\sqrt{s}=7$ TeV, for $E_\perp({\rm lepton})>5$ GeV and for the veto range of $|\eta|<5.2$. The results are presented for different regions of the invariant mass, $M_Y, M_Z$, of the proton dissociation products.}\label{survgam}
\end{center}
\end{table}

To make the interpretation of CEP experimental data less ambiguous it would be helpful to suppress proton dissociation. This can be done by imposing additional veto conditions: no signal in the ZDC and/or, as discussed in the Introduction, FSCs, which cover a larger rapidity interval~\cite{Albrow:2008az,cms}. Another possibility is to select events in which the  centrally produced system has a low transverse momentum. In particular, we can impose a cut on the coplanarity of the $\gamma\gamma$ or $\pi\pi$ pair  in the transverse plane. In the loop formed by two Pomerons between the triple--Pomeron vertex and the matrix element of a `hard' subprocess, $|\mathcal{M}|^2$, the large momentum $p_\perp$ goes through the $|\mathcal{M}|^2$ blob. That is, the centrally produced system  obtains a large transverse momentum which violates the coplanarity of the event. By selecting events with, say, $|\Delta\phi -\pi|<0.1$  we will introduce a cut of $p_\perp < 0.1 E_\perp$\footnote{Recall that the $\Delta\phi$ distribution of CEP diphoton events observed by CDF~\cite{Aaltonen:2011hi}, where the LRG veto is quite good, thus providing a very small admixture of proton dissociation events, has a narrow  peak at $\Delta\phi=\pi$, consistent with the $E_\perp$ of the recoil proton.}.
 
Finally, let us make a few comments concerning the evaluation of the proton dissociation contribution in the case when the central system (e.g. a lepton pair) is produced via the photon--photon fusion mechanism. As a rule (see e.g. \cite{Chatrchyan:2011ci,CMSgam}), an admixture of the processes with proton dissociation is evaluated using the LPAIR event generator \cite{Vermaseren:1982cz,Baranov:1991yq}. However, it is based on purely photon exchanges, and absorption effects caused by the strong interactions between the protons are not accounted for\footnote{We thank Sergey Baranov and Wenbo Li for useful discussions concerning the LPAIR MC.}. This could therefore lead to some overestimation of the admixture caused by proton dissociation, while not affecting the purely exclusive case when the outgoing protons remain intact, where the size of the absorbtive effects are expected to be very small~\cite{Khoze:2000db}. Using a technique similar to that discussed above for the evaluation of the proton dissociative contributions, for example, in the topical case of dilepton production we can evaluate the survival factors $S^2$ (by which the LPAIR predictions should be multiplied). For illustration, in Table~\ref{survgam} we present estimates for the case of 7 TeV proton--proton collisions with $E_\perp({\rm lepton})>5$ GeV, and for the veto range $|\eta|<5.2$.

\section{Meson pair CEP}

\subsection{Symmetric contribution}\label{symc}

\begin{figure}[h]
\begin{center}
\subfigure[]{\includegraphics[clip,trim=0 15 0 0,scale=0.5]{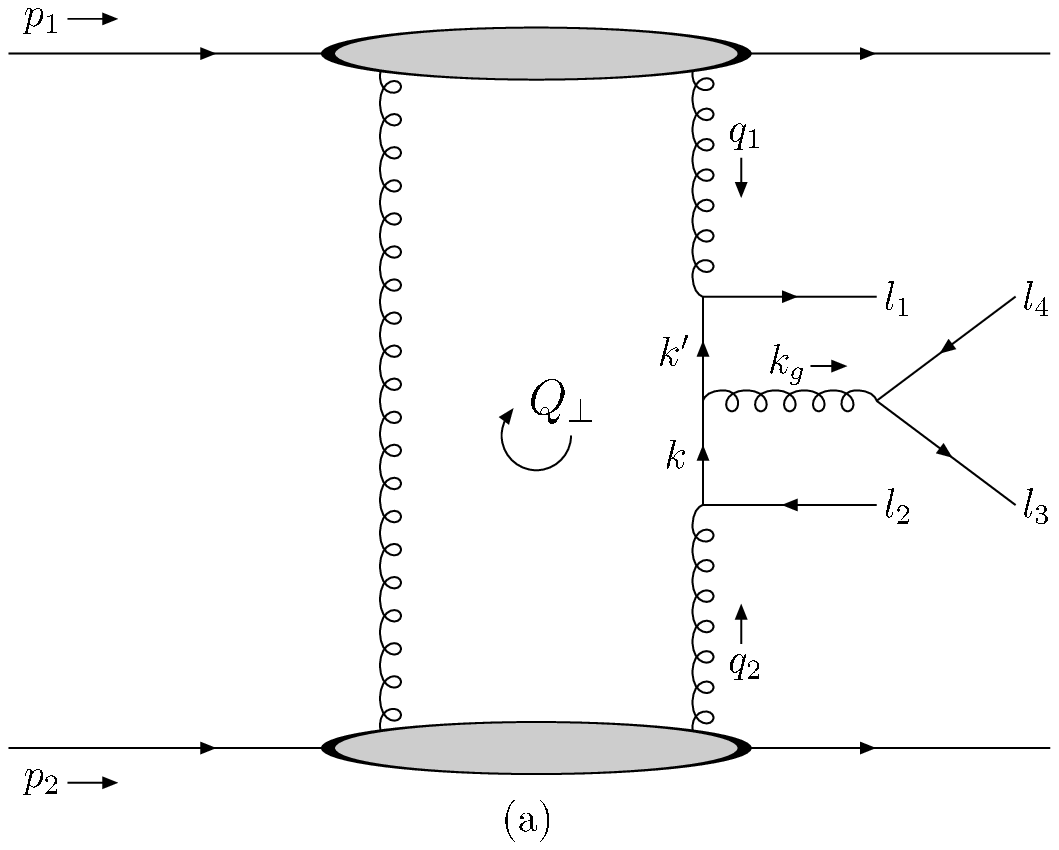}}\qquad
\subfigure[]{\includegraphics[clip,trim=0 15 0 0,scale=0.5]{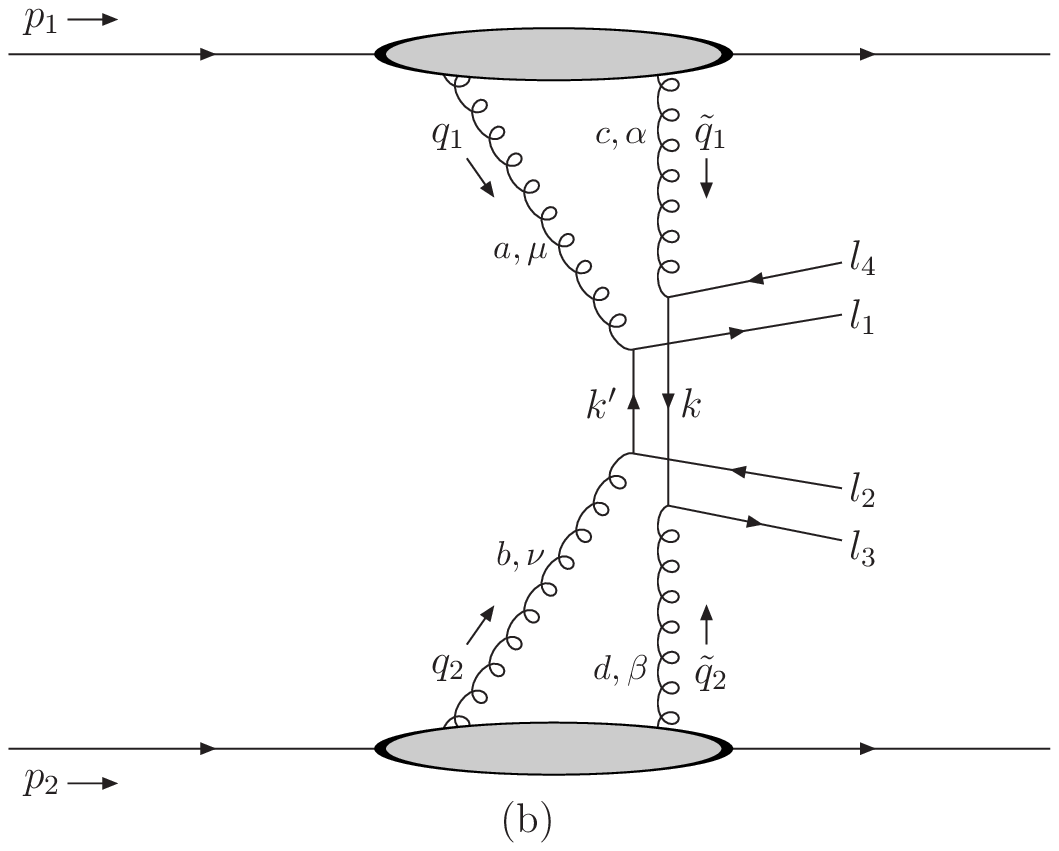}}
\caption{{\bf (a)} Representative perturbative diagram for `skewed' meson pair CEP. {\bf (b)} Representative perturbative diagram for `symmetric' meson pair CEP. In both cases, quarks with momenta $(l_1,l_4)$ and $(l_2,l_3)$ are collinear and form colourless meson states.}\label{fors}
\end{center}
\end{figure}

As discussed in~\cite{HarlandLang:2011qd}, apart from the standard CEP diagram shown in Fig.~\ref{fors}(a), there is a second perturbative mechanism for producing meson pairs at high $k_\perp$. This
is shown in Fig.~\ref{fors}(b): the second $t$--channel gluon now couples directly to a quark line, while the collinear $q\overline{q}$ pairs can be in both flavour--nonsinglet and flavour--singlet combinations. Following~\cite{HarlandLang:2011qd}, we label Figs.~\ref{fors}(a) and (b) as `skewed' and `symmetric' CEP, respectively. In Fig.~\ref{fors}(a) the second $t$--channel gluon is much softer than the fusing gluons, whereas in Fig.~\ref{fors}(b) both $t$--channel gluons participate symmetrically in the hard subprocess. The symmetric diagram represents the perturbative tail to the non--perturbative double Pomeron exchange mechanism, resolving the two--gluon structure of the exchanged Pomerons.

For the calculation of the skewed contribution we use the `hard exclusive' formalism~\cite{Brodsky:1981rp} (see also~\cite{Benayoun:1989ng}), modified to the case of $gg\to M\overline{M}$. The amplitude can be written in the form
\begin{equation}\label{amp}
\mathcal{M}_{\lambda_1\lambda_2}(\hat{s},\theta)=\int_{0}^{1} \,{\rm d}z \,{\rm d}z'\, \phi_M(z)\phi_{\overline{M}}(z')\, T_{\lambda_1\lambda_2}(z,z';\hat{s},\theta)\;,
\end{equation}
where $\hat{s}$ is the $M\overline{M}$ invariant mass squared, $z,z'$ are the meson momentum fractions carried by the quarks, $\lambda_1$, $\lambda_2$ are the gluon helicities and $\theta$ is the scattering angle in the $gg$ c.m.s. frame.
 $T_{\lambda_1\lambda_2}$ is the hard scattering amplitude for the parton level process $gg\to q\overline{q}\,q\overline{q}$, where each (massless) $q\overline{q}$ pair is collinear and has the appropriate colour, spin, and flavour content projected out to form the corresponding meson. In the meson rest frame, the relative motion of the quark and antiquark is small: thus for a meson produced with large momentum, $|\vec{k}|$, we can neglect the transverse component of the quark momentum, $\vec{q}$, with respect to $\vec{k}$, and simply write $q=zk$ in the calculation of $T_{\lambda_1\lambda_2}$. $\phi(z)$ is the meson wavefunction, representing the probability amplitude for finding a valence parton in the meson carrying a longitudinal momentum fraction $z$ of the meson's momentum, integrated up to the scale $Q$ over the quark transverse momentum $\vec{q_\perp}$ (with respect to pion momentum $\vec k$). We then use the formalism outlined in Section~\ref{CEPform} to calculate the $M\overline{M}$ CEP cross section from these subprocess helicity amplitudes, see in particular Section 4.1 (Eq. (41)) of~\cite{HarlandLang:2010ep}.

In~\cite{HarlandLang:2011qd} it was shown that the symmetric amplitude represents a power correction to the standard skewed CEP amplitude $T$, in particular with
\begin{equation}
\frac{T_{\rm sym.}}{T_{\rm skew.}}\sim \frac{\langle Q_\perp^2 \rangle}{k_\perp^2}\;,
\end{equation}
where $k_\perp$ is the meson transverse momentum, and $\langle Q_\perp^2 \rangle$ is the average gluon transverse momentum squared in Fig.~\ref{fors}(a): at high meson $k_\perp$, the symmetric contribution will be subleading. However, at lower values of $k_\perp\sim Q_\perp$ this will not necessarily be the case, and care must be taken to perform an explicit numerical evaluation of the relative contributions of the skewed and symmetric amplitudes.

The symmetric CEP amplitude, considering for simplicity the case of forward outgoing protons ($p_{i_\perp}=0$), can be written in the form

\begin{equation}\label{bsym}
T_{\rm sym.}=\pi^2 \int \frac{d^2 q_{1\perp}}{ q_{1\perp}^4 q_{2\perp}^4 }\,\mathcal{M}_{\rm sym.}\,f_g(x_1,\tilde{x}_1, q_{1_\perp}^2,\mu^2;t_1)f_g(x_2,\tilde{x}_2,q_{2_\perp}^2,\mu^2;t_2) \; ,
\end{equation}
where the notation follows from Fig.~\ref{fors}(b), and this can readily be generalised to the case of non--zero proton $p_\perp$. The subprocess amplitude $\mathcal{M}_{\rm sym.}$ is given by
\begin{align}\label{msym}
\mathcal{M}_{\rm sym.}&=\frac{4}{M_X^4}\frac{1}{N_C^2-1}\delta^{ac}\delta^{bd}q_{1\perp}^\mu q_{2\perp}^\nu q_{1\perp}^\alpha q_{2\perp}^\beta V^{abcd}_{\mu\nu\alpha\beta}\\
&\equiv \frac{4}{M_X^4} V_{\rm sym.}\;.
\end{align}
For example, considering the specific case of scalar non--singlet meson production, as shown in Fig.~\ref{fors}(b), if we identify
\begin{equation}\label{lmom}
l_1=z'k_3 \quad l_2=(1-z)k_4 \quad
l_3=zk_4 \quad l_4=(1-z')k_3\; ,
\end{equation}
then we have
\begin{equation}\label{spec}
V_{\rm sym.} = 16\pi^2 \alpha_S^2\frac{C_F}{2N_C}\int\, {\rm d}z\, {\rm d}z' \,\phi(z)\phi(z')\,\frac{{\rm Tr}({\not}k{\not}q_{2\perp}{\not}l_3
{\not}q_{2\perp}{\not}k'{\not}q_{1\perp}{\not}l_4{\not}q_{1\perp})}{2k^2k'^2}\;.
\end{equation}
The other diagrams corresponding to interchanging the outgoing quark legs $l_1 \leftrightarrow l_2$ and $l_3 \leftrightarrow l_4$, as well as the $s$--channel diagrams containing 3--gluon vertices, should also be included, and the resulting expression can then be combined with (\ref{bsym}) to give an explicit evaluation of the full symmetric CEP amplitude. 

Since we do not know the form of the generalized PDFs, $f_g$, in the kinematic regime relevant to the symmetric CEP contribution the best we can do, following~\cite{HarlandLang:2011qd}, is to put an upper limit on the cross section, based on the Schwarz inequality~\cite{Martin:1997wy}, taking
\begin{equation}
f_g(x,x',Q^2,...)<\frac{1}{2}((f_g(x,-x,Q^2,...)+f_g(x',-x',Q^2,...))\;,
\end{equation}
where the $f_g(x,-x,Q^2,...)$ are the diagonal gluon distribution functions which can be extracted from DIS data. We note that in the symmetric amplitude we must be careful to forbid perturbative emission from both $t$--channel exchanges, which can both be hard, although at a scale that is somewhat lower than $\mu=M_X/2$. In general we can therefore take
\begin{equation}\label{fgsym}
f_g(x,x',Q^2,\mu_1^2,\mu_2^2)=
\frac{1}{2}\frac{\partial}{\partial \log Q^2}\left[\sqrt{T_g(Q^2,\mu_1^2)T_g(Q^2,\mu_2^2)}
(xg(x,Q^2)+x'g(x',Q^2))\right]\;,
\end{equation}
where in the case for example of Fig.~\ref{fors}(b), with the momenta defined as in (\ref{lmom}), we have $\mu_1^2=z'(1-z)M_X^2/4$ and $\mu_2^2=z(1-z')M_X^2/4$. However for simplicity and to give a rough estimate, we set $\mu_1=\mu_2=M_X/4$, corresponding to $\langle z,z' \rangle = 1/2$. It should be noted, however, that in the region of high and low $z,z'$, the `hard' scale can be arbitrarily small, and so the symmetric CEP amplitude may become sensitive to non--perturbative physics. On the other hand, the size of these corrections should be small because the wavefunctions $\phi(z),\phi(z')$ suppress contributions from the the endpoint regions $z,z'=0,1$.

We also note that the expressions for the symmetric amplitude (\ref{msym}) and (\ref{spec}) include a non--negligible contribution from the region where the incoming gluons are far off--mass--shell $Q^2 \sim M_X^2$ in the amplitude evaluation. For example, for the diagram shown in Fig.~\ref{fors}(b), we have
\begin{equation}
\vec{q}_{2\perp}=[z'-(1-z)]\vec{k}_\perp-\vec{q}_{1\perp}\;,
\end{equation}
and therefore $Q_2^2 \sim \vec{q}_{2\perp}^2 \sim \vec{k}_\perp^2$, which is of order the hard scale of the subprocess. In general, the incoming gluon off--shellness, or equivalently $q_\perp$ with respect to the beam direction, is generated by higher order diagrams which describe the creation of a highly virtual gluon from a low--virtuality incoming parton, see the discussion below (\ref{sp}). In the case where the dominant contribution comes from the strongly ordered region ($q_\perp^2 \ll M_X^2$) as in, for example, Higgs CEP, we can factor this initial--state gluon virtuality, generated by these higher order diagrams, into the $q_\perp$--dependent $f_g$'s. As we have $M_X \gg Q_\perp$, corrections due to the off--shellness of the incoming gluons in the $gg \to X$ subprocess will be small, and can be neglected. However, in the situation described above, this is no longer the case and  we should in general consider a full NLO calculation where we explicitly include in the hard process diagrams which generate this high virtuality in the incoming gluons. Such a calculation may include crucial cancellations between, and contributions from, different NLO diagrams which would be missed simply by taking the LO expression as in (\ref{spec}), with the $f_g$'s evaluated at high gluon scales $Q^2 \sim M_X^2$. We can, therefore, only consider (\ref{spec}) as an estimate of the expected rate: in particular the true all--orders cross section may in principle be somewhat larger. 

Such an issue also arises, for example, in the case of skewed $\pi\pi$ CEP, which we recall is strongly suppressed by the $J_z=0$ selection rule due to the fact that the LO on--shell $gg \to \pi\pi$ amplitudes vanish for $J_z=0$ incoming gluons. Here the only non--zero contribution arises from the case when the gluons are off--shell and therefore allow a non--$J_z=0$ component to the cross section: for lower values of $M_X$, a significant contribution can come from the region where the gluon off--shellness is of the same order as the hard scale. The factorised form (\ref{bt}) we use to calculate the $\pi\pi$ cross section, where we ignore the gluon off--shellness in the $gg \to \pi\pi$ amplitudes, can therefore only be considered as an estimate of the full cross section in the lower $M_X$ region: to give a precise evaluation, we would have to calculate the full NLO amplitude, where the off--shellness of the incoming gluons is generated by additional virtual gluons. Na\"{i}vely taking (\ref{bt}) but with the gluon off--shellness explicitly included in the $gg \to \pi\pi$ amplitude will tend to overestimate the cross section, as it will miss the cancellations between different NLO diagrams that we may in general expect to occur\footnote{This raises doubts regarding the consistency of the strategy adopted in~\cite{Lebiedowicz:2011nb,Maciula:2011iv}, based on using explicit results for the amplitudes with off--shell gluons.}, as is observed in the LO $gg \to \pi\pi$ case, leading to the vanishing of the $J_z=0$ amplitude, and the presence of a radiation zero in the $|J_z|=2$ amplitude.

\begin{figure}[h]
\begin{center}
\includegraphics[scale=0.6]{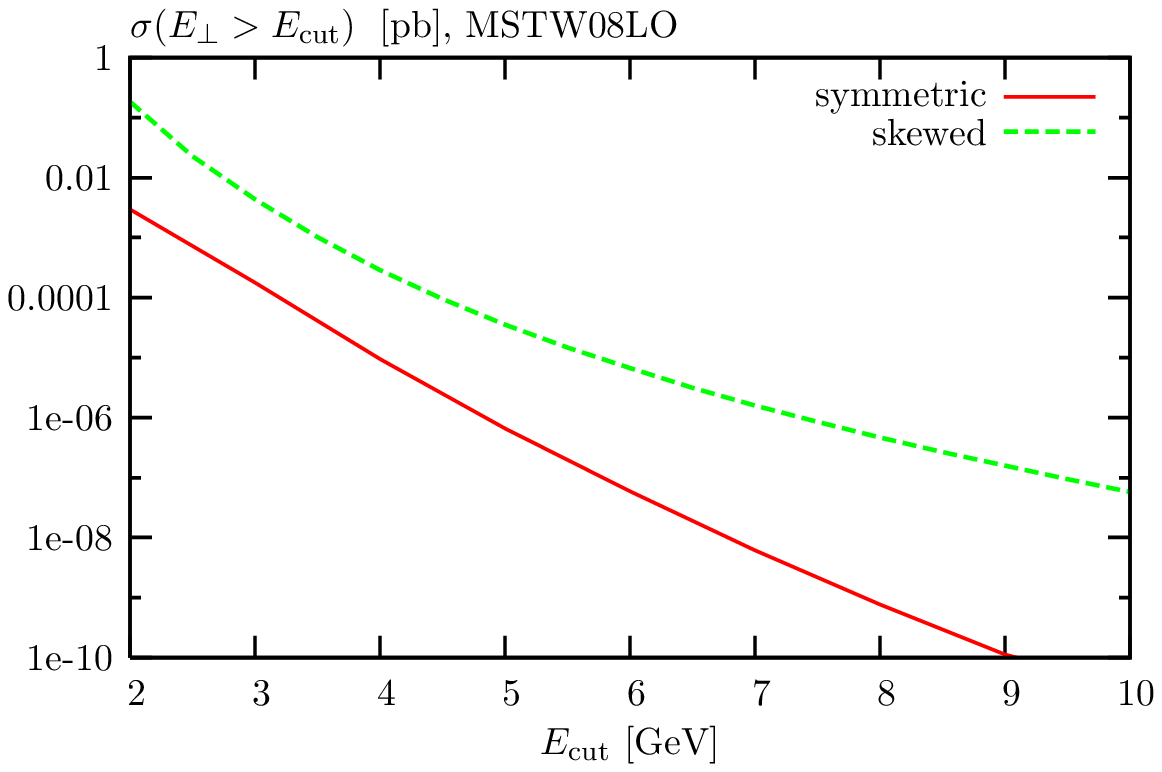}
\includegraphics[scale=0.6]{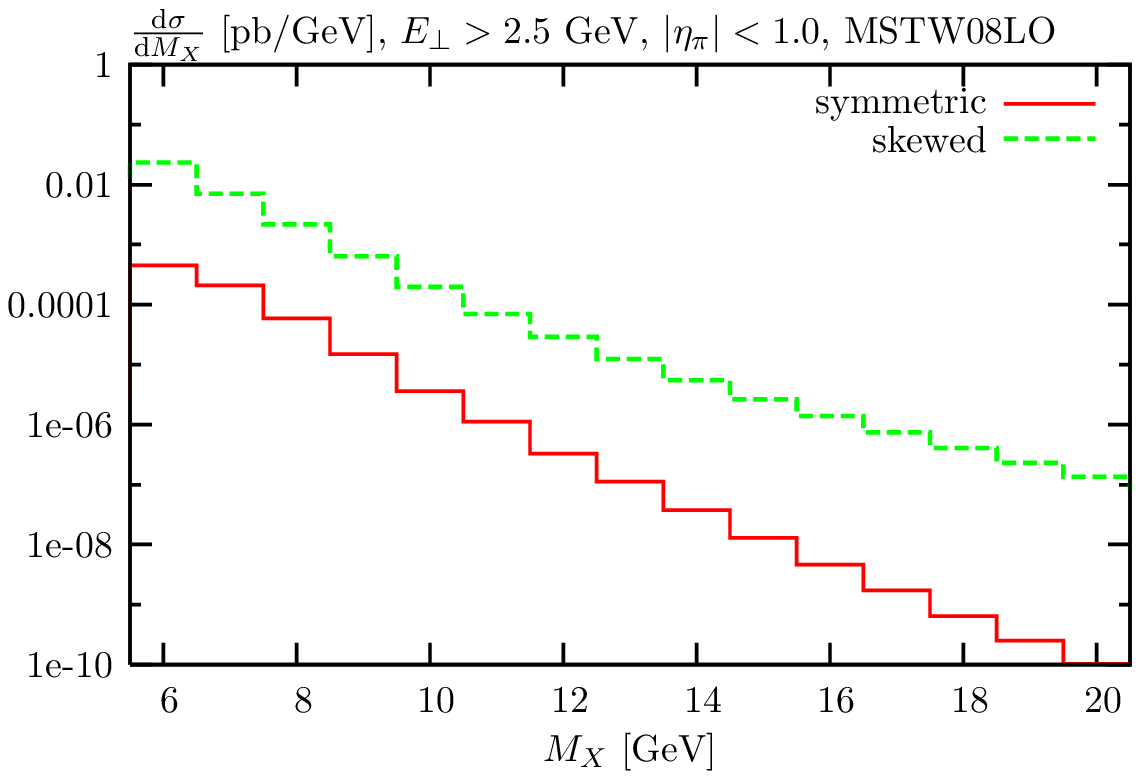}
\caption{Skewed and symmetric contributions to the $\pi^0\pi^0$ CEP cross section for different cuts $E_{\rm cut}$ on the $\pi^0$ transverse energy, $E_\perp$, and the differential cross section ${\rm d}\sigma/{\rm d}M_{\pi\pi}$ for $E_\perp(\pi^0)>2.5$ GeV, at $\sqrt{s}=7$ TeV. In both cases a pseudorapidity cut $|\eta(\pi^0)|<1$ is applied.}\label{sket}
\end{center}
\end{figure}

Finally, using (\ref{bsym}) and (\ref{fgsym}) we can, for example, calculate the symmetric contribution to $\pi^0\pi^0$ CEP. This is plotted along with the standard skewed $\pi^0\pi^0$ CEP cross section in Fig.~\ref{sket}, as a function of the cut on pion $E_\perp$. Also shown is the differential cross section ${\rm d}\sigma/{\rm d}M_{\pi\pi}$. We can see that, even at lower values of the pion $E_\perp$, the symmetric contribution is expected to be very small, and well within other theoretical uncertainties on the skewed cross section prediction. As we would expect, recalling that the symmetric contribution represents a power correction to the skewed contribution, as $E_\perp$ increases this suppression becomes stronger. We have therefore explicitly shown that the skewed $\pi^0\pi^0$ contribution is negligible in the entire kinematic region, and this will be equally true for the CEP of other flavour non--singlet states ($\pi^+\pi^-$, $K^+K^-$...). Moreover, as (\ref{fgsym}) gives an upper bound on the $f_g$'s, the symmetric cross section may in general be even smaller than our numerical estimate suggests (although it is also true that a proper treatment of the gluon off--shellness may increase the cross section). Recalling that the $\pi\pi$ skewed CEP cross section is already strongly reduced by the $J_z=0$ selection rule, this relative suppression of the symmetric CEP contribution will be even stronger for other meson pair production processes ($\eta'\eta'$, $\eta\eta'$...) which have non--zero $J_z=0$ production cross sections at LO. We can therefore safely neglect this symmetric contribution when it comes to producing numerical estimates for the expected meson pair CEP cross sections in Section~\ref{mres}.

\subsection{Non--perturbative process}\label{nps}
\begin{figure}[h]
\begin{center}
\includegraphics[scale=0.9]{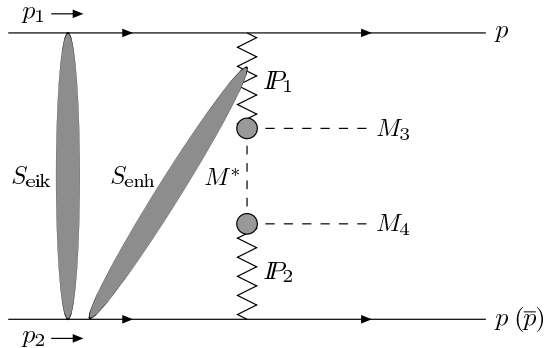}
\caption{Representative diagram for the non--perturbative meson pair ($M_3$, $M_4$) CEP mechanism, where $M^*$ is an intermediate off--shell meson of type $M$. Eikonal and (an example of) enhanced screening effects are indicated by the shaded areas.}\label{npip}
\end{center}
\end{figure}
The formalism of Section~\ref{CEPform} is only valid for relatively large values of the central system mass $M_X$: in the low mass (low $k_\perp$) region, we expect a non--perturbative picture of the type shown in Fig.~\ref{npip} to give the dominant contribution. For this one--meson--exchange mechanism (see for instance~\cite{Pumplin:1976dm,Azimov:1974fa,Lebiedowicz:2011nb}), the $M\overline{M}$ pair is created via double--Pomeron exchange, with an intermediate $t$--channel off--shell meson, and the amplitude is calculated using the tools of Regge theory, see~\cite{HarlandLang:2011qd} for more details\footnote{This subject has been also addressed in \cite{Lebiedowicz:2011nb}, although we have strong reservations concerning their treatment of the final state interaction between the outgoing mesons, for the reasons described above (\ref{spipi}).}. In general, we should include both these contributions to the total meson pair CEP cross section (recalling from~\cite{HarlandLang:2011qd} that there is only a small interference between the dominantly imaginary perturbative and the dominantly real non--perturbative amplitudes), although in the limit of low/high meson $k_\perp$ only the perturbative/non--perturbative descriptions will be applicable.

\begin{figure}[h]
\begin{center}
\includegraphics[scale=0.7]{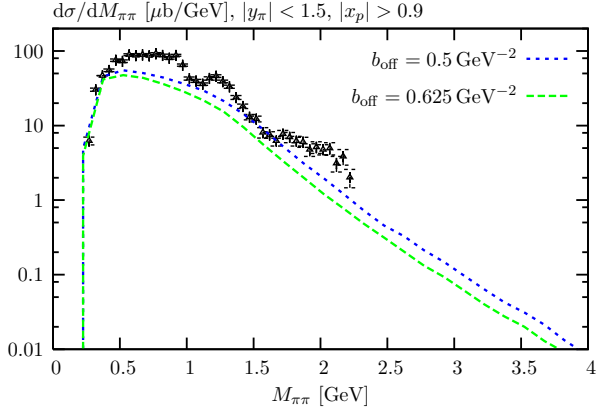}
\caption{Differential cross section ${\rm d}\sigma/{\rm d}M_{\pi\pi}$ for the non--perturbative contribution to $\pi^+\pi^-$ CEP at $\sqrt{s}=62$ GeV, compared to CERN--ISR data~\cite{Breakstone:1990at}. The theory curves are calculated using the model described in the text, and with secondary Reggeons included, using the fit of~\cite{Lebiedowicz:2009pj}. In all cases, the pions are restricted to lie in the rapidity region $|y_\pi|<1.5$ and the cut $|x_p|>0.9$ is imposed on the outgoing protons.}\label{ISR}
\end{center}
\end{figure}

To calculate the non--perturbative $M\overline{M}$ contribution, we note that the CEP cross section is given by
\begin{equation}\label{ncross}
\sigma^{CEP}=\frac {S^2}{16\pi(16\pi^2)^2}\int dp^2_{1\perp}dp^2_{2\perp}dy_1dy_2dk^2_{\perp}\frac{|\mathcal{M}|^2}{s^2}\;,
\end{equation}
where $\sqrt{s}$ is the c.m.s. energy, $p_{1\perp}, p_{2\perp}$ are transverse momenta of the outgoing protons, $k_\perp$ is the meson transverse momentum and $y_i$ are the meson rapidities. $S^2$ is the soft survival factor -- see below for a discussion of this. 
The matrix element is given by $\mathcal{M}=\mathcal{M}_{\hat{t}}+\mathcal{M}_{\hat{u}}$, with $\hat{t}=(P_1-k_3)^2$, $\hat{u}=(P_1-k_4)^2$, where $P_i$ is the momentum transfer through Pomeron $i$, and $k_{3,4}$ are the meson momenta. We have
\begin{equation}\label{namp}
\mathcal{M}_{\hat{t}}=\frac 1{M^2-\hat{t}} F_p(p^2_{1\perp})F_p(p^2_{2\perp})F^2_M(\hat{t})\sigma_0^2
\bigg(\frac{s_{13}}{s_0}\bigg)^{\alpha_P(0)}\bigg(\frac{s_{24}}{s_0}\bigg)^{\alpha_P(0)}\;,
\end{equation}
where $M$ is the meson mass and we take $s_0=1\,{\rm GeV}^2$ and $\alpha_P(p^2_{i\perp})=1.08-0.25|p^2_{i\perp}|$, for $p^2_{i\perp}$ measured in ${\rm GeV}^2$~\cite{Donnachie:1992ny}, and $s_{ij}=(p_i'+k_j)^2$ is the c.m.s. energy squared of the final state proton--meson system $(ij)$. The proton form factors are as usual taken to have an exponential form, $F_p(t_i)=\exp(B_it_i/2)$, while the slope of the Pomeron trajectory is included in the definition of the slope\footnote{We note that in the equivalent formula to (\ref{namp}) in~\cite{HarlandLang:2011qd}, the factors of $\alpha(p^2_{i\perp})$ should read $\alpha_P(0)$.},
\begin{equation}\label{bpom}
B_i=b_0+2\alpha'\log\bigg(\frac{s_{ij}}{s_0}\bigg)\;,
\end{equation}
with $b_0=4\,\,{\rm GeV}^{-2}$.  Concentrating on the case of $\pi\pi$ production, the overall cross section normalisation is set by the total pion--proton cross section $\sigma(\pi p)=\sigma_0 (s_{ij}/s_0)^{\alpha(0)-1}\approx30$ mb at the relevant sub--energy for $\pi\pi$ production at the LHC. The factor $F_M(\hat{t})$ in (\ref{ncross}) is the form factor of the intermediate off--shell meson and, as discussed in~\cite{HarlandLang:2011qd}, it is quite poorly known, in particular for larger values of $\hat{t}$. It seems reasonable to take a typical `soft' exponential form $F_M(\hat{t})=\exp{(b_{\rm off}(\hat{t}-M^2)})$, and the value of the slope can be approximately fitted to reproduce the correct normalisation of CERN--ISR data~\cite{Breakstone:1990at}, as shown in Fig.~\ref{ISR}, where a fair fit\footnote{The contribution of secondary Reggeons is also included here, using the fit of~\cite{Lebiedowicz:2009pj}. We are grateful to Piotr Lebiedowicz for providing a scan of the data found in~\cite{Breakstone:1990at}.} to the data is given by the choice $b_{\rm off}=0.5 \,{\rm GeV}^{-2}$. While this value, which we note corresponds to quite a high average meson $k_\perp^2\sim -\hat{t}$ for a typical `soft' process, reproduces the overall data normalisation, we note that the phase space region ($M_{\pi\pi}\lesssim 2$ GeV, $|y_\pi|<1.5$) in which the data are collected places a very limited constraint on the cross section contribution with relatively large pion $k_\perp$, and so the higher $\hat{t}$ behaviour of $F_M(\hat{t})$, and therefore the non--perturbative cross section (which is sensitive to $\sim F_M^4$), is still quite uncertain. With this is in mind we also show the prediction using the lower value of $b_{\rm off}=0.625 \,{\rm GeV}^{-2}$, which we will also use when making numerical predictions in Section~\ref{mres}. While this seems to give too small a cross section compared to the CERN--ISR data, nonetheless we consider this value to indicate the sort of spread in predictions we may expect at higher pion $k_\perp$, where the existing data place little constraint on the form of $F_\pi$.

It is worth emphasising that the theory curves shown in Fig.~\ref{ISR} should be considered only as a low (non--resonant) limit of the actual experimental cross section.  The region $M_{\pi\pi}<2$ GeV manifests a number of
resonances ($f_2(1270), f_0(1370), f_0(1500), f_2'(1525), f_2(1950),...$) which overlap with each other. Besides this, the $t$-channel state $M^*$ could correspond not only to pion exchange but also to the exchange of heavier states ($a_1,\ a_2,...$), which may enlarge the CEP cross section and explain the rather low value of the slope $b_{\rm off}$ used to describe the data in Fig.~\ref{ISR} (see also~\cite{Lebiedowicz:2011nb}). Recall also that within the quasi--eikonal formalism the probability of higher mass pion excitation is about 0.45~\cite{Kaydalov:1971ta}, which is rather large, compared to 0.15 for the proton dissociation case, indicating a non--negligible contribution
from high mass meson, $M^*$, states\footnote{On the other hand the effect of low mass {\em proton} dissociation, which is included in the models of~\cite{Ryskin:2011qe,Ryskin:2009tk} tends to reduce the survival probability, $S^2_{\rm eik}$, and therefore the exclusive cross section. In Fig.~\ref{ISR} we use these models to calculate the `eikonal' survival factor, $S^2_{\rm eik}$. While there is tension between these models and the latest TOTEM data on elastic $pp$ scattering at 7 TeV~\cite{Antchev:2011zz}, with the models in particular giving a probability of low mass proton dissociation which is smaller than that indicated by the data, the updated model discussed in Section~\ref{mres}, which accounts for this, gives a cross section normalisation that is too low to fit the ISR data for realistic values of $b_{\rm off}$. We therefore treat the results of the one--pion--exchange model with this higher value of $S^2_{\rm eik}$ as giving a phenomenological description to the data, but note that a more complete treatment, including for example the exchange of other higher mass resonances ($a_1,a_2$,...) may be required to fully describe the ISR data. It is also not completely excluded that the higher value of $S^2_{\rm eik}$ that the previous models of~\cite{Ryskin:2011qe,Ryskin:2009tk} give may be more suitable, see Section~\ref{mres} for further discussion.}.

\begin{figure}[h]
\begin{center}
\includegraphics[scale=0.7]{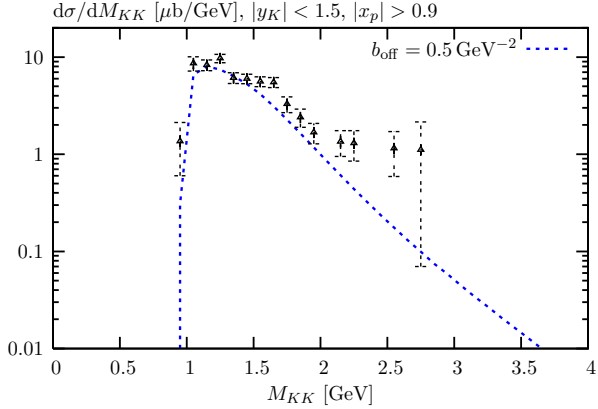}
\caption{Differential cross section ${\rm d}\sigma/{\rm d}M_{KK}$ for the non--perturbative contribution to $K^+K^-$ CEP at $\sqrt{s}=62$ GeV, compared to CERN--ISR data~\cite{Breakstone:1989ty}. The theory curves are calculated using the model described in the text, and with secondary Reggeons included, using the fit of~\cite{Lebiedowicz:2011tp}. In all cases, the kaons are restricted to lie in the rapidity region $|y_K|<1.5$ and the cut $|x_p|>0.9$ is imposed on the outgoing protons.}\label{KISR}
\end{center}
\end{figure}

In Fig.~\ref{KISR} we show the prediction using the same non--perturbative model for $K^+K^-$ CEP, compared to ISR data~\cite{Breakstone:1989ty}. Far enough above the $KK$ threshold, we expect the $K^+K^-$ differential cross section ${\rm d}\sigma/{\rm d}M_{KK}$ to be roughly suppressed by a factor\footnote{This observation is in strong contrast to the much higher level ($\sim 2$ orders of magnitude at $M_{\pi\pi(KK)}=4$ GeV) of suppression predicted in~\cite{Lebiedowicz:2011tp}, see e.g. Fig.~7 (Right, dashed line), compared to Fig.~5 (Right, solid line) of~\cite{Lebiedowicz:2009pj} (Figure numbers refer to published versions), which to our understanding are calculated using a very similar model to the one we describe here, with these plots in particular corresponding to the situation prior to the inclusion of $\pi\pi$($KK$) rescattering, which may in principle complicate the issue. We can find no reason for a such an extreme level of suppression, and indeed Fig.~7 of~\cite{Lebiedowicz:2011tp} appears to be in tension with the ISR data~\cite{Breakstone:1989ty}. Although contact was made with the authors to clarify this, no clear justification for this (very large) discrepency was given by them.} of $(\sigma_{K p}/\sigma_{\pi p})^4 \sim 1/2$ (at ISR energies this is complicated by the non--negligible contribution of secondary Reggeons, although the level of suppression is still roughly the same). There will be some additional suppression from the Kaon mass $m_K>m_\pi$ nearer the $KK$ threshold, and in particular the integrated $K^+K^-$ cross section will be signficantly suppressed due to this threshold effect. 

Finally, we have to include an additional suppression factor to calculate the genuinely exclusive $M\overline{M}$ ($=\pi\pi$,...) cross section, i.e. that due to screening corrections. This was discussed in~\cite{HarlandLang:2011qd}, and here we give some clarifying remarks. We are considering the screening corrections to the non--perturbative amplitude shown in Fig.~\ref{npip}; in terms of the Reggeon formalism these corrections are described by the exchange of additional (one or more) Pomerons.  First, there is the exchange between the two incoming (outgoing) protons ($p_1,p_2$), which is just the usual eikonal survival factor $S_{\rm eik}$. Next, we have to consider the exchange between the upper (lower) proton and the lower (upper) meson, that is between $p_1$ ($p_2$) and $M_4$ ($M_3$). We do not consider the exchange between $p_1$ ($p_2$) and $M_3$ ($M_4$) as for the corresponding pion--proton amplitude we have used the Donnachie--Landshoff parametrisation~\cite{Donnachie:1992ny}, in which the {\em effective} Pomeron $P_1$ ($P_2$) exchange already includes the sum of eikonal--like multi--(bare) Pomeron diagrams. Together with the possible exchange between the upper (lower) proton and the lower (upper) Pomeron, this corresponds to the {\em enhanced} screening, $S_{\rm enh}$, shown in Fig.~\ref{npip}. Note however that due to the rather small phenomenological value of the triple--Pomeron coupling (see for instance~\cite{Kaidalov:1973tc,Kaidalov:1979jz,Luna:2008pp}) the main effect is expected to be from the secondary proton--meson interaction. However, in this case at higher values of $\sqrt{\hat{s}}$ the size of the screening effect will be strongly suppressed by the small size of the produced pions $\sim 1/\sqrt{\hat{s}}$. In the perturbative approach, this screening effect is therefore power suppressed by the small absorptive cross section $\sim 1/k_\perp^2$ of the meson in the early stage of its formation\footnote{To understand this we should recall that in the early production stage (in particular, in the case of a large $t$-channel meson $M^*$ virtuality) the meson colour field does not have sufficient time to regenerate, and the meson exists as a {\it half-dressed} (see e.g.~\cite {Dokshitzer:1988bq,Dokshitzer:1991wu}) particle (or pre-hadron, see~\cite{Kopeliovich:2011zza} for a recent review) with transverse size (much) smaller than that of the fully formed hadrons. Therefore, its interaction cross section should be weaker than in the case of an asymptotic (completely hadronized) meson state.}, and can be safely ignored in the phase space region we will consider here. In the nonperturbative case, unfortunately, there is no clear prescription for treating this rescattering of secondaries, as the cross section is written in terms of `effective' Pomeron trajectories which are already renormalized by absorptive effects. As we do not know the details of how this renormalization occurs, it is therefore difficult to evaluate the level of proton--meson rescattering that we will expect to occur. Nonetheless, for reasonable values of the meson $k_\perp$ which we will be considering we expect the size of this rescattering to be similarly suppressed in the non-perturbative case, and we will ignore such an effect in our numerical calculations.

	Next, we recall that there is no secondary interaction between two outgoing mesons mediated by Reggeon exchange. This can be seen as follows: the time, $\tau$, needed to form a Reggeon increases linearly with energy, $\tau\propto E$, and at high meson--meson energy, $E \sim \sqrt{s_{M\overline{M}}}$, the two mesons travel apart at very close to the velocity of light. Since the meson pair production time in Fig.~\ref{npip} is practically instantaneous ($\sim 1/E$), while a much longer time ($\sim E/m^2$) is need for the formation of a Reggeon by the secondary meson, there is insufficient time for a Reggeon emitted by one meson to interact with the other\footnote{We thank Yuri Dokshitzer for discussions about this space--time argument.}. In other words, the two outcoming high energy mesons cannot talk to each other. Formally, when we consider the diagram with Reggeon exchange between the two outgoing mesons, with the Reggeon represented by a ladder--type graph, and check the imaginary part, $i\epsilon$, of the propagators, we see that the two (or three) poles which give the leading contribution at large $s_{M\overline{M}}$ are all on the same side of the integration contour, and so closing the contour on the other side gives a zero leading $s_{M\overline{M}}$ contribution. As was shown by Mandelstam \cite{Mandelstam:1963cw}, the leading $s$ contribution in the case of additional Reggeon exchange only comes from non--planar diagrams and not from planar graphs, of the type discussed in~\cite{Amati:1962nv}. Thus the only possibility to have rescattering of the objects at high $s_{M\overline{M}}$ without violating causality is to consider the rescattering of incoming pomerons $P_1$ and $P_2$, but this is suppressed numerically by the small value of the triple--Pomeron vertex (see for instance~\cite{Kaidalov:1973tc,Kaidalov:1979jz,Luna:2008pp}).  We do not account for such an effect here. Instead, following~\cite{HarlandLang:2011qd}, we introduce an extra suppression of the form of $\exp(-n)$, corresponding to the small Poisson probability not to emit other secondaries in the $I\!\!P I\!\!P\to M_3\overline{M}_4$ process at the initial meson pair production stage (rather than being due to final--state interactions between the mesons). Here $n(s_{M\overline{M}})$ is the mean number of secondaries. This factor may be described as the Reggeization of the $M^*$ meson exchange, which means that we now deal with non--local meson--Pomeron vertices and the $t$--channel meson $M^*$ becomes a non--local object, i.e. it has its own size. It is this non--locality that is responsible for the non--violation of causality. More precisely for the case of $\pi\pi$ CEP we take 
	\begin{align}\nonumber
	n(s_{\pi\pi}) &=0 \qquad  &M_{\pi\pi}<M_{f_2(1270)}\;,\\ \label{spipi}
	n(s_{\pi\pi}) &=c\ln \left(\frac{M_{\pi\pi}^2}{s_0}\right) \qquad  &M_{\pi\pi} \geq M_{f_2(1270)}\;,
	\end{align}
	with $s_0=M_{f_2(1270)}^2$ and $c=0.7$-- these precise values are taken for definiteness, but we note that in either case different choices are certainly possible. As discussed above, this choice appears to give a fair description of the existing ISR data. In this way, we account for the fact that we expect no additional suppression in the lower mass resonance region, where additional interactions at the meson pair production state lead mainly to the
	formation of resonances and not to the production of new secondaries.
	%, and the suppression is a function the free energy, $E_{\rm free}=M_{\pi\pi}-2M_\pi$, available for the creation of secondary particles. 
	A similar although slightly modified procedure is taken for the case of $K^+K^-$ CEP. In particular we replace $M_{f_2(1270)}$ in (\ref{spipi}) with $M_{f_2(1525)}$, and account for the fact that we should expect the number of secondaries $n(s_{KK})$ to be a function of the free energy, $E_{\rm free}=M_{KK}-2M_K$, available for the creation of secondary particles, which can be numerically important because of the larger Kaon mass $M_K$. 

\section{Results: predictions for the LHC, RHIC and the Tevatron.}

\subsection{Meson pair CEP}\label{mres}

In this Section we present some numerical predictions for meson pair CEP at the LHC and RHIC. For the sake of brevity, we only consider the case of $\pi\pi$ CEP, although similar prediction for, say, $KK$ CEP can readily be made (see Section~\ref{nps}). For the perturbative contribution, i.e. that found using the formalism outlined in Section~\ref{CEPform}, we use the MSTW08L0 PDFs, as these give a $\gamma\gamma$ CEP cross section that is in better agreement with the recent CDF data (see Section~\ref{gamnlo}). For the non--perturbative contribution, we use the formalism described in Section~\ref{nps}. To calculate the soft survival factor we use the models of~\cite{Ryskin:2011qe,Ryskin:2009tk}, incorporating new results for $S^2_{\rm eik}$ using the most recent fit~\cite{Ryskin:2012ry}, which includes the TOTEM data on elastic $pp$ scattering at 7 TeV~\cite{Antchev:2011zz}. As the previous models of soft diffraction~\cite{Ryskin:2011qe,Ryskin:2009tk} gave a somewhat lower value for $\sigma_{\rm tot}$ at 7 TeV than the value measured by TOTEM, this most recent fit requires an increased optical density $\Omega(b_t)$ of the proton, in particular at low impact parameter, $b_t$, see~\cite{Ryskin:2012ry} for more details. Because of this increased optical density the probability of additional soft interactions is increased and the survival factor $S^2_{\rm eik}$ calculated using this model is lower, giving a factor of $\sim 2$ decrease in the the combined $S^2=S^2_{\rm eik}S^2_{\rm enh}$ for $\gamma\gamma$ CEP at $\sqrt{s}=7$ TeV, when compared with the previous predictions in~\cite{HarlandLang:2010ep,HarlandLang:2010ys,HarlandLang:2010te,HarlandLang:2011qd,HarlandLang:2011zz}. However, at this stage it is possible that with further theoretical and/or experimental developments (for example, early measurements of diffractive dijet production by CMS~\cite{HollarECT} indicate that the gap survival probability may be higher than that given by our updated model), this value of $S^2_{\rm eik}$ may increase, and so the following $\gamma\gamma$ CEP predictions should be viewed as slightly conservative with respect to the gap survival probability. 

On the other hand, we find that for example the $\pi\pi$ CEP cross section is almost unchanged by taking this updated model for $S^2_{\rm eik}$. We recall that the new optical density $\Omega(b_t)$, motivated by the TOTEM data~\cite{Antchev:2011zz}, increases mainly at low $b_t$. On the other hand the LO $gg\to \pi\pi$ amplitude vanishes for $J_z=0$ gluons, and so $\pi\pi$ CEP only occurs at LO for $|J_z|=2$ fusing gluons. This contribution, due to its spin structure, $b_{t\mu}b_{t\nu}$, in impact parameter space, vanishes as $b_t\to 0$. Therefore the more peripheral $\pi\pi$ CEP cross section is less sensitive (in comparison with $\gamma\gamma$ CEP where the $J_z=0$ contribution dominates) to the increase in $\Omega(b_t)$ at low $b_t$: this is equally true for other CEP process ($\chi_{c(1,2)}$, $K^+K^-$...) that vanish in the forward proton limit.

\begin{figure}
\begin{center}
\includegraphics[scale=0.7]{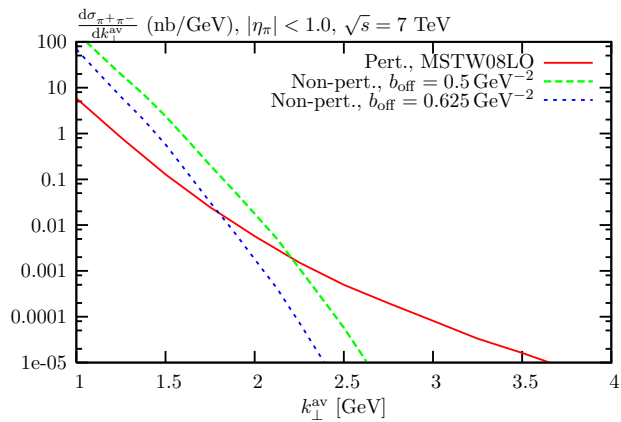}
\caption{Perturbative and non--perturbative contributions to non--resonant $\pi^+\pi^-$ CEP differential cross sections ${\rm d}\sigma/{\rm d}k_\perp^{\rm av}$ at $\sqrt{s}=7$ TeV, where $k_\perp^{\rm av}=(k_\perp(\pi^+)+k_\perp(\pi^-))/2$ is the scalar average of the pion's transverse momentum, for the acceptance region $k_\perp>0.5$ GeV and $|\eta_\pi|<1.0$.}\label{ptmplots}
\end{center}
\end{figure}

\begin{table}
\begin{center}
\begin{tabular}{|l|c|c|c|}
\hline
&$(\eta_{\rm min},\eta_{\rm max})$&$k_\perp({\rm min})$($|\vec{k}|({\rm min})$) [GeV]&$\sigma^{\rm Non-pert.}$ [nb]\\
\hline
ALICE&$(-0.9,0.9)$&0&840--1100  \\
\hline
ATLAS&$(-2.5,2.5)$&0.5&1100--1700\\
\hline
CMS&$(-2,2)$&0.2&5300--6700\\
\hline
LHCb&$(2,5)$&3&3200--4000\\
\hline
STAR&(-1.0,1.0)&0.15&2500--3100\\
\hline
\end{tabular}
\caption{Pseudorapidity ($\eta$) ranges and minimum $k_\perp$ values corresponding to the regions with good detection efficiency for $\pi^+\pi^-$ CEP for the LHC detectors and the STAR detector at RHIC. In the case of the ALICE detector, an additional cut, $M_{\pi\pi}> 0.8$ GeV, is imposed, and in the case of LHCb the momentum cut is placed on $|\vec{k}|({\rm min})$ and not $k_\perp$. Also shown is the (non--perturbative) non--resonant $\pi^+\pi^-$ CEP cross section, in nb, within these acceptance regions. The upper (lower) values of the cross sections correspond to taking $b_{\rm off}=0.5(0.625)\,{\rm GeV}^{-2}$, as described in Section~\ref{nps}.}\label{acc}
\end{center}
\end{table}

\begin{table}[h]
\begin{center}
\begin{tabular}{|l|c|c|c|c|}
\cline{2-3}\cline{4-5}
\multicolumn{1}{c|}{}&\multicolumn{2}{c|}{$k_\perp>1.5$ GeV}&\multicolumn{2}{c|}{$k_\perp>2$ GeV}\\
\hline
&Non--pert.&Pert.&Non--pert.&Pert.\\
\hline
ALICE&23--120&14&0.058--0.69&0.82\\
\hline
ATLAS&170--720&50&0.76--7.0&3.0\\
\hline
CMS&130--580&47&0.53--5.2&2.9\\
\hline
LHCb&69--310&10&0.23--2.2&0.53\\
\hline
STAR&39--190&1.1&0.11--1.2&0.067\\
\hline
STAR ($p_{\perp i}<0.4$ GeV)&13-76&0.31&0.025-0.39&0.018\\
\hline
STAR ($p_{\perp i}>0.5$ GeV)&6.1-23&0.15&0.026-0.21&0.008\\
\hline
\end{tabular}
\caption{Perturbative and non--perturbative contributions to non--resonant $\pi^+\pi^-$ CEP cross sections (in pb) at the LHC ($\sqrt{s}=7$ TeV), for acceptance regions defined in Table~\ref{acc}, but with an additional cut on the pion $k_\perp$, as indicated in the table. Also shown are the cross sections at RHIC energies ($\sqrt{s}=500$ GeV), with and without cuts on the outgoing proton transverse momenta $p_{\perp i}$. The upper (lower) values of the non--perturbative cross sections correspond to taking $b_{\rm off}=0.5(0.625)\,{\rm GeV}^{-2}$, as described in Section~\ref{nps} and in the text.}\label{cs23}
\end{center}
\end{table}

In Table~\ref{acc} we show the cuts that we will impose to roughly account for the pseudorapidity and momentum acceptance of the LHC detectors and the STAR detector at RHIC for meson pair CEP~\cite{exper}. We recall that the STAR detector is currently expected to have the ability to tag protons and therefore select purely exclusive events. At the LHC, without forward proton detectors exclusivity must instead be selected by vetoing on additional hadronic activity in a large enough rapidity region, although in this case there will also be some contribution from non--exclusive events, in particular where one or both of the protons dissociate, as in (\ref{dd}). Such a background contribution could be particularly important at LHCb, where the veto region ($2<\eta<5$ and $-4<\eta<-2$ for charged tracks) is fairly limited, see Section~\ref{dcep} for further discussion. However, as discussed in the Introduction, recently the possibility of measuring exclusive low mass pion pair production using the ALFA RP detectors during the special low pile--up LHC runs has been proposed~\cite{Staszewski:2011bg}, and data taken as part of a proposed dedicated CMS run of single no--pile--up interactions, engaged and taking data in common with TOTEM would be especially promising~\cite{cms-totem}.

Using the cuts of Table~\ref{acc} and taking $\pi^+\pi^-$ as a specific example\footnote{We recall that the $\pi^0\pi^0$ CEP cross--section prediction is $1/2$ of the $\pi^+\pi^-$ cross section, from isotopic invariance and the identity of the final state $\pi^0$ states.}, we then also show in Table~\ref{acc} the non--resonant $\pi^+\pi^-$ CEP cross section. As discussed in Section~\ref{nps}, the size of the non--perturbative cross section at larger values of the meson $k_\perp$ is largely uncertain, due to our lack of knowledge of the form factor of the intermediate off--shell pion $F_\pi$. We therefore show approximate upper (lower) bounds on the range of predictions, found using the values for the slope parameter $b_{\rm off}=0.5 (0.625)\,{\rm GeV}^{-2}$, assuming an exponential functional form $F_\pi(\hat{t})=\exp{(b_{\rm off}(\hat{t}-M^2)})$. In this case the predicted cross section, which is primarily sensitive to lower values of the meson $k_\perp$, does not change too much between these possibilities, although as we shall see this is no longer true when $k_\perp$ cuts are imposed. We only show the cross--section prediction calculated within the non--perturbative framework, because at the low $M_{\pi\pi}$ and $k_\perp(\pi)$ values accessible due to the reasonably low $k_\perp$ reach of the current LHC experiments and the relatively large rapidity coverage, the application of the pQCD approach used to describe both the overall CEP process, as well as the $gg\to \pi^+\pi^-$ subprocess cross section, cannot be justified. In particular, it gives an unphysical divergent answer as the scale of $\alpha_S$ and the off--shellness of the propagators in the $gg\to \pi^+\pi^-$ subprocess decrease.

We can see in Table~\ref{acc} that the expected $\pi^+\pi^-$ cross sections, calculated within the non--perturbative framework, are quite large. This can even be the case as we increase $M_X$, when the pions are produced at non--zero pseudorapidity and can still have fairly low $k_\perp$. However, in Fig.~\ref{ptmplots}, we show the differential cross section ${\rm d}\sigma/{\rm d}k_{\perp}^{\rm av}$ (where $k_{\perp}^{\rm av}$ is a suitably defined transverse momentum variable, see the Figure caption) in both the perturbative and non--perturbative case, for the two choices of $b_{\rm off}$ described above. Clearly there is a strong damping in the non--perturbative cross section at higher $k_\perp$, caused by the soft pion form factor $F_\pi(k_\perp^2)$, which drops sharply with the pion $k_\perp$: in this region, the high $k_\perp$ tail of the $\pi^+\pi^-$ CEP process is generated by a purely pQCD mechanism, which cannot be predicted within the framework of Regge theory. However we can see that the different choices of $b_{\rm off}$, while predicting fairly similar total cross sections (see Table~\ref{acc}), give roughly an order of magnitude variation in the differential cross sections at higher $k_\perp$. Furthermore, as discussed in Section~\ref{nps}, the existing CERN--ISR data~\cite{Breakstone:1990at} imposes some constraint on the off--shell pion form factor, $F_\pi$, at low but not high pion $k_\perp$, and so the non--perturbative cross section may in principle lie outside this range (although for such a `soft' process we would not expect the slope parameter to be much smaller than the lower choice $b_{\rm off}=0.5 \, {\rm GeV}^{-2}$); these estimates represent our best educated extrapolation to the higher $k_\perp$ region. 

Nonetheless, if we impose suitable cuts on the pion $k_\perp$, the perturbative contribution should dominate and can be used to reliably estimate the full $\pi^+\pi^-$ CEP cross section. In Table~\ref{cs23} we show cross--section predictions for the same experimental pseudorapidity regions, but with the additional cut on the pion $k_\perp>2.0$ GeV, and the perturbative mechanism is expected to be non-negligible and possibly dominant. For higher values of the $k_\perp$ cut (see Fig.~\ref{ptmplots}) we therefore expect the non-perturbative contribution to be negligible, although we can see in Table~\ref{acc} that this is in a region where the perturbative cross section ($\sim$ 1 pb or lower) is quite small and unfortunately may not be experimentally accessible for the low luminosity runs suitable for CEP observations. This is due to the strong ($\sim$ 1/100) suppression by the $J_z=0$ selection rule of the perturbative $\pi^+\pi^-$ CEP cross section\footnote{Recalling the observed enhancement by LHCb~\cite{LHCb} of the $\chi_{c2}$ rate above theory expectations, which may be due to a significant contamination from non--exclusive events, we note that the observed $\pi^+\pi^-$ rate may be correspondingly enhanced by such contamination at the LHC, where the outgoing protons are not tagged.} that comes from the fact that the $gg\to \pi^+\pi^-$ amplitude vanishes at LO for $J_z=0$ incoming gluons~\cite{HarlandLang:2011qd}.  We also show in Table~\ref{cs23} the $\pi\pi$ CEP cross section with the cut $k_\perp>1.5$ GeV: in this case, while the cross sections are higher, the non--perturbative contribution is expected to be dominant. As noted in~\cite{HarlandLang:2011qd}, other related processes, such as $\eta(')\eta(')$ CEP, where the production amplitudes do not vanish for $J_z=0$ incoming gluons, are expected to have significantly larger cross sections and so may represent experimentally more realistic and theoretically cleaner (that is, with a smaller non--perturbative contribution) observables. 

\begin{figure}[h]
\begin{center}
\includegraphics[scale=0.6]{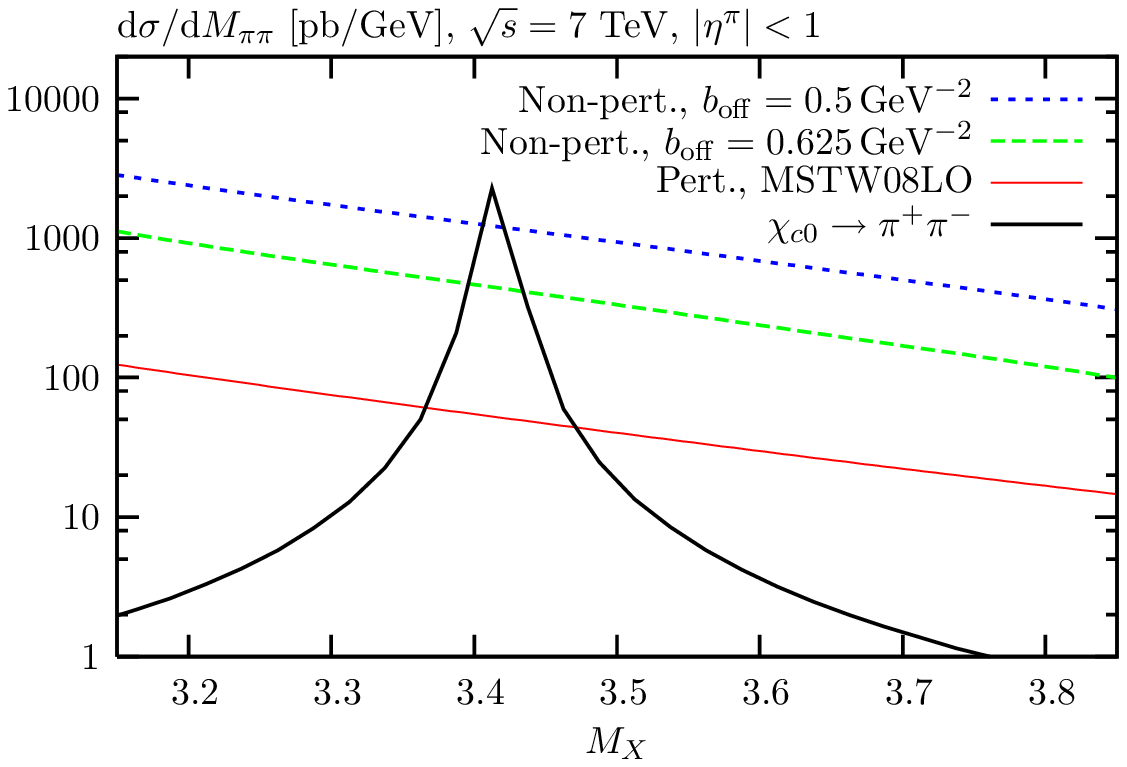}
\includegraphics[scale=0.6]{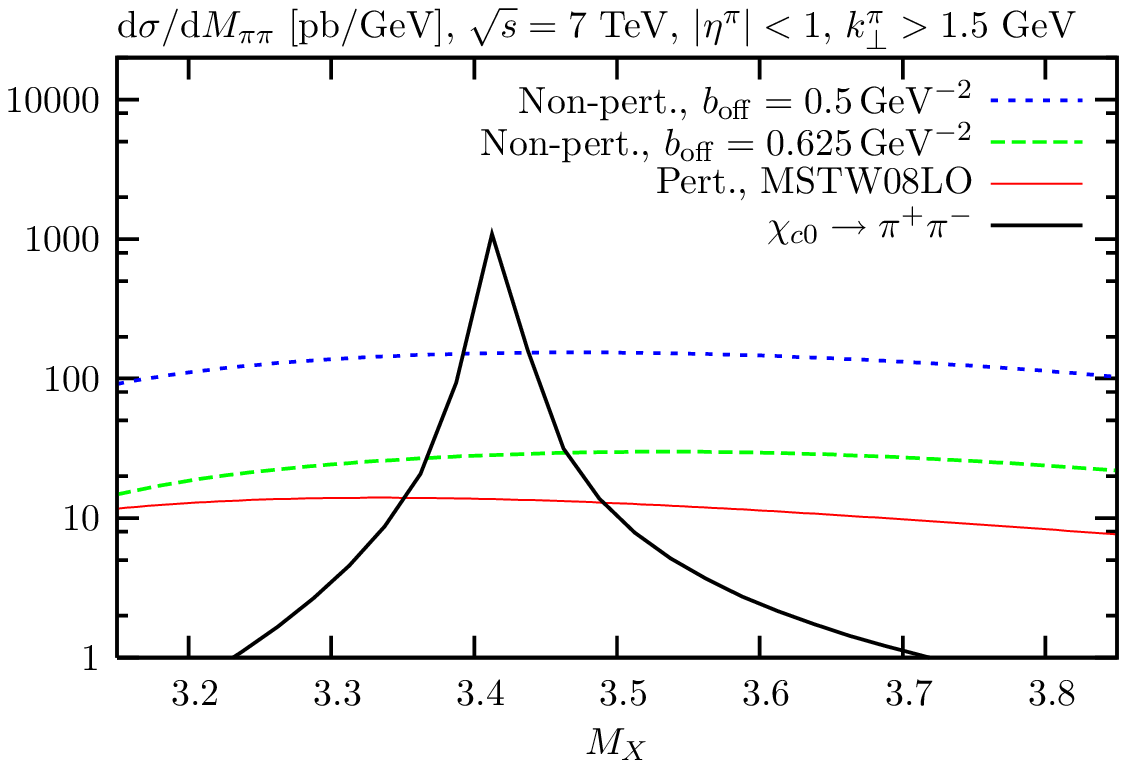}
\caption{$\chi_{c0} \to \pi^+\pi^-$ Breit--Wigner peak, and perturbative and non--perturbative (for different $b_{\rm off}$ values) contributions to non--resonant $\pi^+\pi^-$ CEP in the $\chi_{c0}$ mass region at $\sqrt{s}=7$ TeV, for different $k_\perp$ and $\eta$ cuts on the pions.}\label{lhcm}
\end{center}
\end{figure}

\begin{figure}[h]
\begin{center}
\includegraphics[scale=0.6]{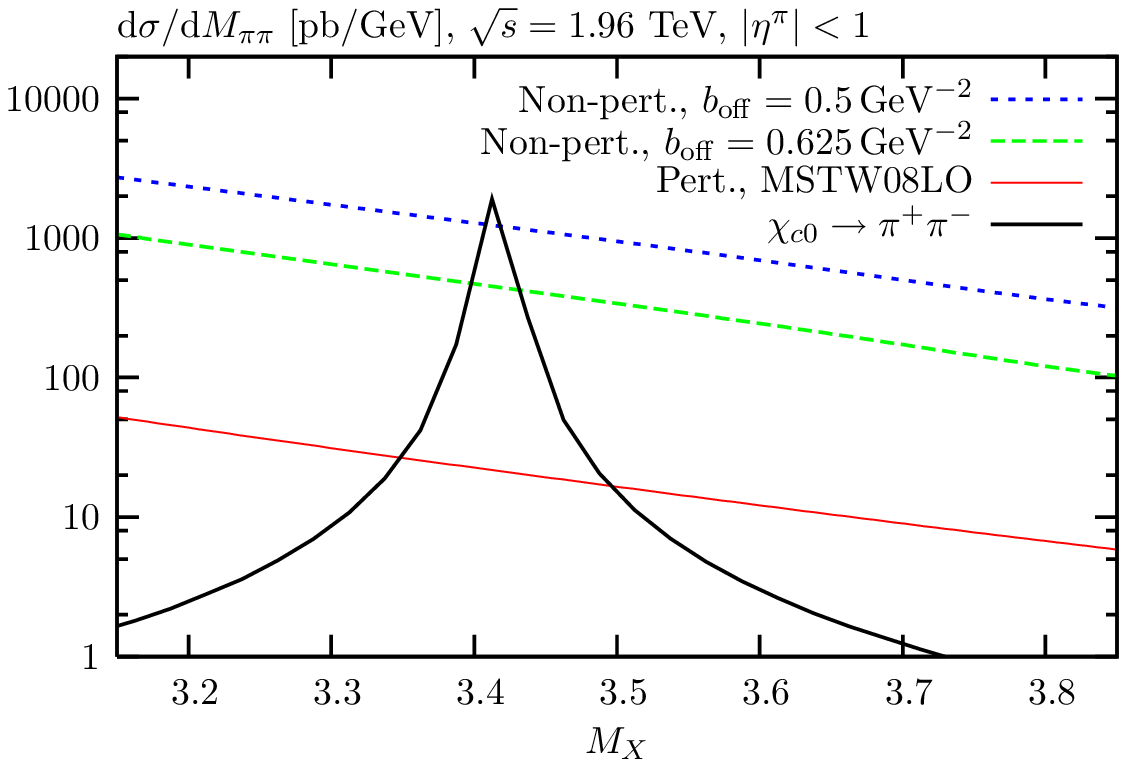}
\includegraphics[scale=0.6]{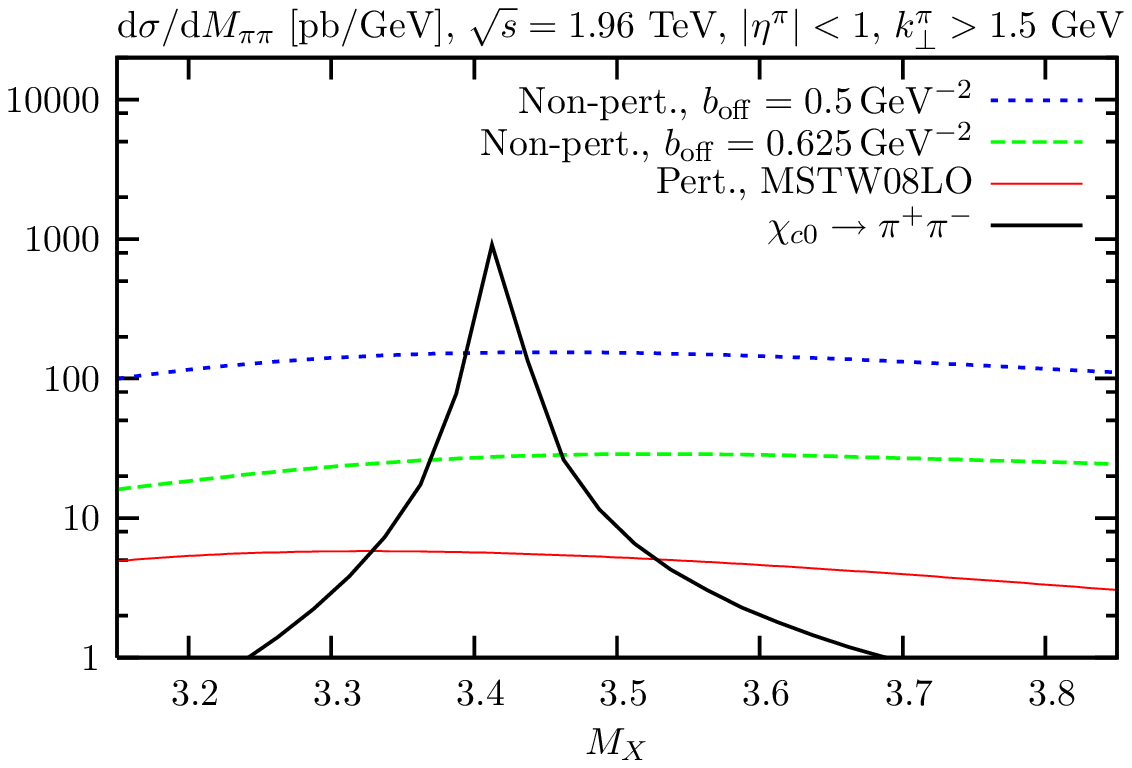}
\caption{$\chi_{c0} \to \pi^+\pi^-$ Breit--Wigner peak, and perturbative and non--perturbative (for different $b_{\rm off}$ values) contributions to non--resonant $\pi^+\pi^-$ CEP in the $\chi_{c0}$ mass region at $\sqrt{s}=1.96$ TeV, for different $k_\perp$ and $\eta$ cuts on the pions.}\label{tevm}
\end{center}
\end{figure}

\begin{figure}[h]
\begin{center}
\includegraphics[scale=0.6]{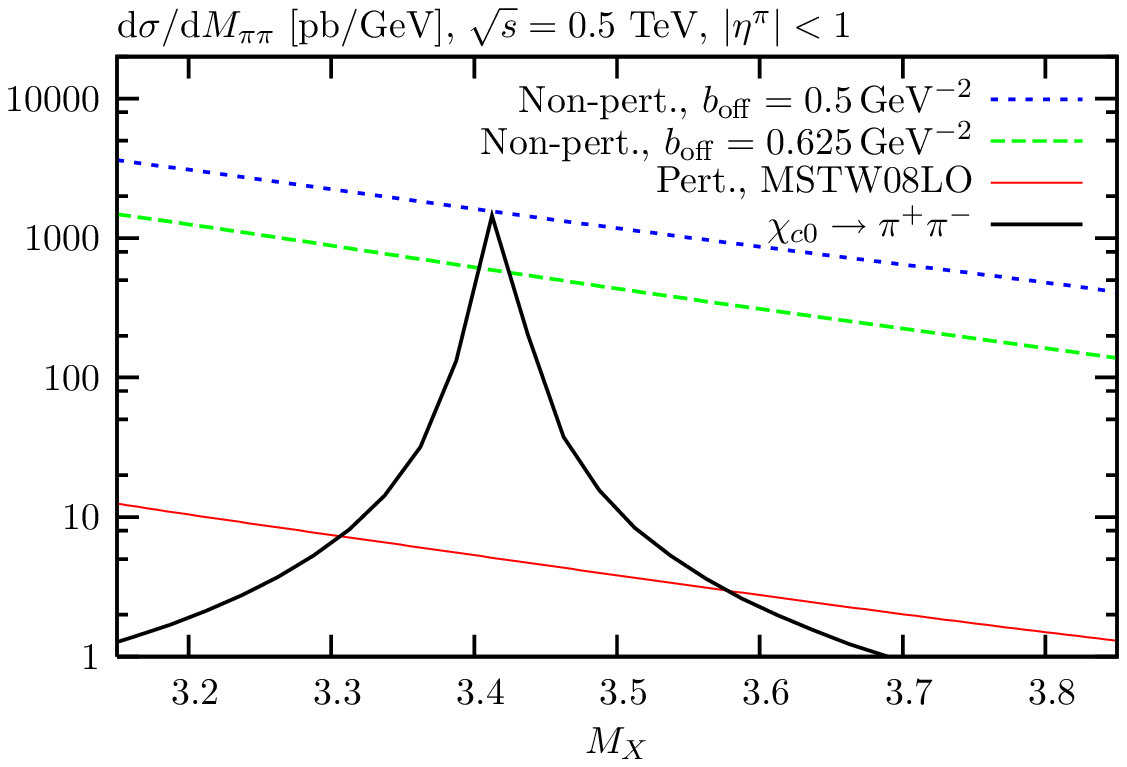}
\includegraphics[scale=0.6]{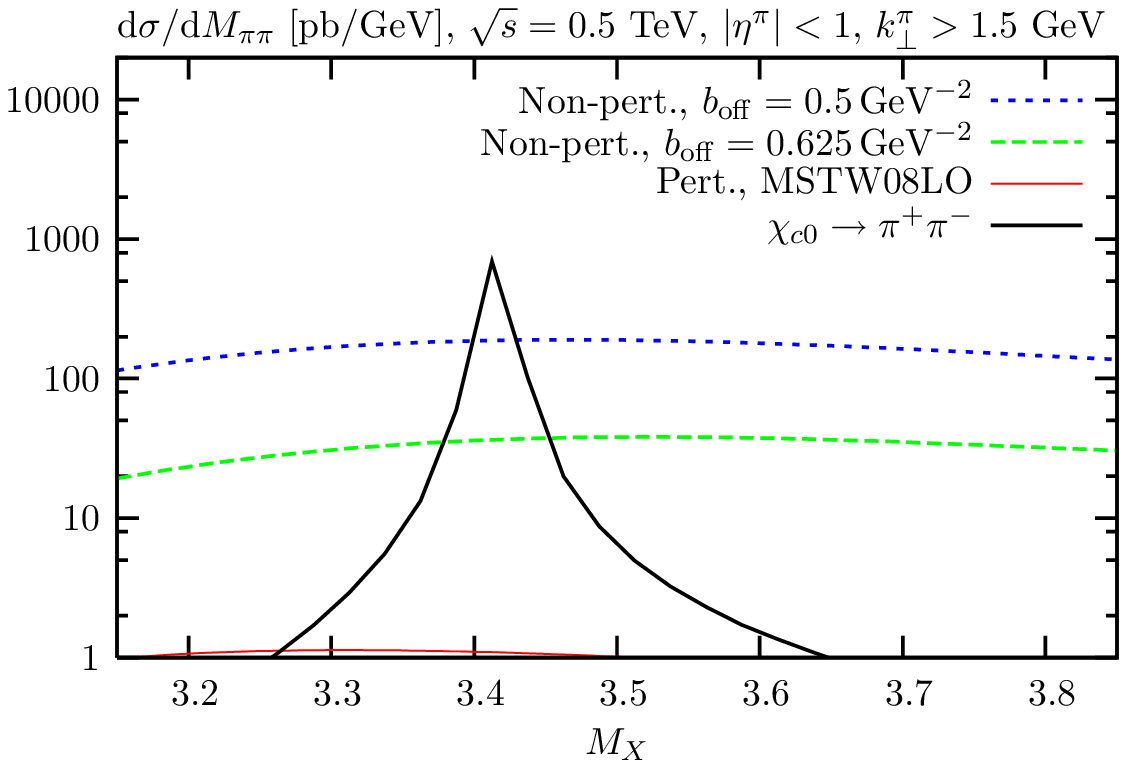}
\caption{$\chi_{c0} \to \pi^+\pi^-$ Breit--Wigner peak, and perturbative and non--perturbative (for different $b_{\rm off}$ values) contributions to non--resonant $\pi^+\pi^-$ CEP in the $\chi_{c0}$ mass region at $\sqrt{s}=0.5$ TeV, for different $k_\perp$ and $\eta$ cuts on the pions.}\label{rhicm}
\end{center}
\end{figure}

We recall that at the STAR detector at RHIC, a large CEP data sample with tagged forward protons is expected to be available in the near future~\cite{Guryn:2008ky,Lee:2010zzp,Guryn:2011zz}. The position of the existing and proposed proton tagging detectors is ideally suited for the observation of meson pair CEP in the experimentally accessible mass region. In Table~\ref{cs23} we also show the $\pi^+\pi^-$ cross section after acceptance cuts on the proton transverse momenta $p_{\perp i}$ have been imposed, with in particular the first cut ($p_{\perp i}<0.4$ GeV) corresponding to the RPs currently installed (Phase I) and the second cut ($p_{\perp i}>0.5$ GeV) to the phase II RPs, to be installed in $\sim 2013$. Such an evaluation requires an explicit inclusion of the proton ${\bf p}_\perp$ distributions {\it after} the inclusion of soft survival effects. We can see that the relative perturbative to non--perturbative contribution is expect to be much lower than at $\sqrt{s}=7$ TeV, which is due to the different $\sqrt{s}$ dependence of the production mechanisms: while the perturbative contribution drops quite steeply with  $\sqrt{s}$ due to the fairly steep $x$--dependence of the gluon density, $g(x,Q^2)$, the non--perturbative contribution exhibits a less steep, Regge--like energy dependence, see (\ref{namp}), and may even increase with decreasing $\sqrt{s}$ due to the larger survival factor $S^2_{\rm eik}$. It therefore appears that at the STAR detector the perturbative contribution will be negligible, although we note that this mechanism, which will dominantly occur for $|J_z|=2$ pions, will exhibit a different distribution in $\phi$, the azimuthal angle between the outgoing protons~\cite{HarlandLang:2010ep}, than in the non--perturbative case, where no such restriction on the pion angular momentum is present. The observation of $\pi^+\pi^-$ CEP with tagged forward protons would therefore represent an interesting test of these two mechanisms and their relative importance.

\begin{table}[h]
\begin{center}
\begin{tabular}{|l|c|c|c|}
\hline
&$k_\perp>2$ GeV&$k_\perp>2.5$ GeV&$k_\perp>5.5$ GeV\\
\hline
$\sqrt{s}=1.96$ TeV ($|\eta|<1$)&9.0&3.1&0.059\\
\hline
$\sqrt{s}=7$ TeV ($|\eta|<2.4$)&19&6.1&0.12\\
\hline
\end{tabular}
\caption{Perturbative cross section ratio $\sigma^{\rm pert.}(\pi^+\pi^-)/\sigma(\gamma\gamma)$ (in \%) at different c.m.s. energies and for different cuts on the pion/photon $k_\perp$ and $\eta$. MSTW08LO PDFs are used, with the survival factors $S^2_{\rm eik}$, $S^2_{\rm enh}$ calculated as described in the text.}\label{cs4}
\end{center}
\end{table}

In Table~\ref{cs4} we show the (perturbative) cross section ratio $\sigma^{\rm pert.}(\pi^+\pi^-)/\sigma(\gamma\gamma)$ for different cuts on the pion/photon $k_\perp(\pi,\gamma)$ at the LHC ($\sqrt{s}=7$ TeV): for such an observable, the theoretical uncertainties from the choice of PDF (and to a lesser extent those due to the survival factors $S_{\rm enh}^2$, $S_{\rm eik}^2$) practically cancel out, and such ratios therefore provide a better test of the underlying theory. As we can see, for $k_\perp>2$ GeV the $\pi^+\pi^-$ cross section is predicted to be a factor of $\sim 5$ smaller than the $\gamma\gamma$ cross sections. Any excess above this might be evidence for a non--negligible non--perturbative $\pi\pi$ contribution, which Table~\ref{cs23} indicates may be present; for the higher $k_\perp$ cuts in Table~\ref{cs4} such a contribution can on the other hand be safely ignored, see Fig.~\ref{ptmplots}. As $k_\perp$ increases the $\pi^+\pi^-$ cross section is strongly suppressed, which is due to the factor of $(f_\pi/k_\perp)^4$ present in the $gg \to \pi^+\pi^-$ cross section, representing the small probability for each outgoing $q\overline{q}$ pair to form a pion. We recall that the cross section ratios shown in Table~\ref{cs4} are calculated using the new results for $S^2_{\rm eik}$ from the most recent fit~\cite{Ryskin:2012ry}, which includes the TOTEM data on elastic $pp$ scattering at 7 TeV~\cite{Antchev:2011zz}, and results in a factor of $\sim 2$ decrease in the $\gamma\gamma$ cross section, while leaving the more peripheral $\pi\pi$ cross section roughly unchanged. However, as described in the beginning of this Section, it is possible that this lower value of $S^2_{\rm eik}$ is somewhat conservative, and so the cross section ratios shown in Table~\ref{cs4} may be slightly lower. For comparison, we also show the same cross section ratio $\sigma^{\rm pert.}(\pi^+\pi^-)/\sigma(\gamma\gamma)$ at the Tevatron, but using the previous (higher) values of $S^2$ calculated using the models described in~\cite{Ryskin:2011qe,Ryskin:2009tk}, and used in for example~\cite{HarlandLang:2010ys,HarlandLang:2011qd}, see Section~\ref{pdfcomp} for further discussion.

Finally, in Figs.~\ref{lhcm}--\ref{rhicm} we show results for the $\pi^+\pi^-$ CEP cross section in the $\chi_{c0}$ mass region at LHC, Tevatron and RHIC energies, which will be relevant for evaluating the potential continuum background to resonant $\chi_{c0}\to\pi^+\pi^-$ production. The $\pi^+\pi^-$ mass distribution from $\chi_{c0}$ decay is given by a simple non--relativistic Breit--Wigner, with the $\chi_{c0}$ cross section normalisation set using the \texttt{SuperCHIC} MC~\cite{SuperCHIC}, which implements the theory described in~\cite{HarlandLang:2010ep}. We can see that once basic $\eta$ cuts are imposed on the final state pions, the $\chi_{c0}$ signal is expected to lie at a similar level to the non--perturbative background with $b_{\rm off}=0.5 \, {\rm GeV}^{-2}$, and that the perturbative contribution is expected to be negligible. Although we note that the non-perturbative background may be somewhat lower in this mass region (for comparison we also show the background for the choice $b_{\rm off}=0.625 \, {\rm GeV}^{-2}$, which gives a lower cross section), it is not completely clear that the signal peak will be visible over the background, taking into account the various theory uncertainties (experimental resolution effects may also decrease the S/B ratio). However, if we also impose a simple $k_\perp>1.5$ GeV cut on the final--state pions, we can see that the background is strongly reduced with a much smaller effect on the $\chi_c$ signal rate (for which the $\chi_c$ mass $M_\chi \approx 3.5$ GeV ensures that a large fraction of the central pions have $k_\perp>1.5$ GeV): the predicted $\chi_{c0}\to\pi^+\pi^-$ rate lies (at least) an order of magnitude above the expected background. We can therefore safely conclude that even within the (in principle quite large) theory uncertainties, $\chi_{c0}\to \pi^+\pi^-$ is expected to represent a clean experimental signal, with a low continuum background once suitable cuts are imposed.

\subsection{$\gamma\gamma$ CEP: PDF comparison}\label{pdfcomp}

\begin{table}[h]
\begin{center}
\begin{tabular}{|l|c|c|c|c|c|c|}
\hline
&\footnotesize{MSTW08LO}&\footnotesize{CTEQ6L}&\footnotesize{GJR08LO}&\footnotesize{MRST99}&\footnotesize{CT10}&\footnotesize{NNPDF2.1}\\
\hline
$\sqrt{s}=1.96$ TeV ($|\eta|<1$) &1.4&2.2&3.6&0.35&0.47&0.29\\
\hline
$\sqrt{s}=7$ TeV ($|\eta|<1$)&0.061&0.069&0.16&0.013&0.0094&0.0057\\
\hline
$\sqrt{s}=7$ TeV ($|\eta|<2.5$)&0.18&0.20&0.45&0.039&0.027&0.017\\
\hline
\end{tabular}
\caption{$\gamma\gamma$ CEP cross sections (in pb) for different choices of gluon PDF, at $\sqrt{s}=1.96$ and 7 TeV, and for different cuts on the photon pseudorapidity, $\eta$. The photons are restricted to have transverse energy $E_\perp>2.5$ GeV at $\sqrt{s}=1.96$ TeV and $E_\perp>5.5$ GeV at $\sqrt{s}=7$ TeV.}\label{pdft}
\end{center}
\end{table}

As discussed in Section~\ref{gamrev}, even after the various theoretical (and experimental) uncertainties are taken into account, there is some tension between the recent CDF $\gamma\gamma$ CEP data~\cite{Aaltonen:2011hi} and the theoretical prediction using the NLO MRST99 PDF set, while if the MSTW08LO PDF set is taken there is good agreement between data and theory. Although some caution is needed, recalling the theoretical uncertainties, we can in principle use these CEP measurements to shed some light on the gluon PDF in this low--$x$ and low--$Q^2$ region, where it is poorly determined. We recall in particular the significant difference between the LO and NLO PDF fits: while the LO PDFs have quite a steep low--$x$ dependence, the NLO PDFs are much smaller and can even be negative, at small $x$ and $Q^2$ for the more modern fits. With this in mind, in Table~\ref{pdft} we show cross section predictions for $\gamma\gamma$ CEP within the CDF kinematics, at $\sqrt{s}=1.96$ TeV, and we can see that for a representative sample of LO PDF sets, there is good agreement with the data, while the NLO set predictions (we are careful to choose only those sets which are {\it not} negative in the relevant kinematic region) all lie significantly below it. Future LHC measurements would allow us to begin to study the energy dependence of the $\gamma\gamma$ cross section, and through this we may hope to provide further information about the gluon PDF, in particular at even lower values of $x$. In Table~\ref{pdft}, we therefore also show predictions for the LHC ($\sqrt{s}=7$ TeV) with the same PDF sets, and for different cuts on the photon pseudorapidity, $\eta$. 

We note that the values in the lowest row ($|\eta|<2.5$) are somewhat lower than those given in Fig.~7 of~\cite{CMSgam}, which is principally due to the different (lower) value for $S^2_{\rm eik}$ we now use to account for the recent TOTEM data~\cite{Antchev:2011zz}, as well as various other theoretical refinements (the `enhanced' survival factor, $S^2_{\rm enh}$, non-zero proton $p_\perp$ etc) not included in the \texttt{ExHuME} MC~\cite{Monk:2005ji} that is used to calculate these numbers\footnote{We should also recall that, as discussed in Section~\ref{dcep}, accounting for the possibility of proton dissociation would lead to an increase of the theoretical CEP prediction by about a factor of 2, which is consistent with the CMS upper limit.}. On the other hand, as discussed at the beginning of Section~\ref{mres}, we note that the value of $S^2_{\rm eik}$ may in principle be higher for $\gamma\gamma$ CEP, in which case the LHC cross section predictions in Table~\ref{pdft} will also be correspondingly increased. With this in mind, for the Tevatron predictions we keep the previous (higher) values of $S^2$ calculated using the models described in~\cite{Ryskin:2011qe,Ryskin:2009tk}, and used in for example~\cite{HarlandLang:2010ys,HarlandLang:2011qd}, consistently with the cross section predictions quoted in~\cite{Aaltonen:2011hi}. However, we note that using the value of $S^2_{\rm eik}$ ($S^2_{\rm enh}$) calculated using the fit of~\cite{Ryskin:2012ry} the corresponding $\gamma\gamma$ cross section prediction will be a factor of $\sim 2$ lower, although this appears to be disfavoured by the CDF data.

\section{Conclusions}

We have performed a detailed theoretical and phenomenological study of several benchmark central exclusive production processes in high--energy proton--(anti)proton collisions, focusing in particular on $\gamma\gamma$ and meson pair CEP at the Tevatron, RHIC and the LHC. Our approach is based on the successful and well--motivated Durham perturbative model, but we have also addressed the importance of additional non--perturbative contributions in the case of meson pair CEP, which are expected to be important for lower meson $k_\perp$. Unlike the case of inclusive production of massive particles and final states, even in the perturbative sector there are large uncertainties in the predictions, in particular from unknown higher order perturbative corrections, from the choice of gluon PDFs and from the soft survival factors.\footnote{In practice it is only approximately true that we can separate these effects, since the emission of additional gluons can contribute to all three.}

Our current understanding of these uncertainties can be summarized as follows. 
\begin{itemize}

\item[(i)]
The dependence on the gluon PDF is amplified by the fact that the CEP cross section is essentially proportional to $(xg(x))^4$. We have argued that it is reasonable to take the range spanned by leading order and next--to--leading order PDFs as indicative of the uncertainty, and have used MRST08 LO and MRST99 NLO for this purpose. We noted that the former (which has a higher density of small--$x$ gluons)  gives better agreement with the recent CDF diphoton data~\cite{Aaltonen:2011hi}. We emphasise again that at the low $x$ values that are relevant for the cases of charmonium, diphoton and dimeson CEP, the difference between the LO and NLO PDFs can be very large.

\item[(ii)] Our CEP `hard' matrix elements are calculated at leading order in QCD perturbation theory. NLO (and higher order) corrections come from a number of sources, of which the most important are:

\begin{itemize}

\item[(a)] NLO corrections to the hard $gg\to X$ matrix element.

\item[(b)] NLO effects in the unintegrated parton densities, see~\cite{Martin:2009ii}.

\item[(c)] Effects of radiation off the screening gluon, of the type discussed in~\cite{Khoze:2009er}.

\item[(d)] A systematic account of self--energy insertions in the propagator of the screening gluon\footnote{Usually in Feynman diagrams we should include a wavefunction renormalization factor $Z^{1/2}$ for each external parton leg. However the conventional PDFs are defined in a different way, with the whole $Z$ factor (for each of the two legs) included in the PDF, while the `hard' matrix element is calculated by taking $Z=1$ for each of the external parton legs. For standard inclusive processes this prescription leads to the correct result, but some care is needed in the CEP case. Since the soft screening gluon in Fig.~\ref{fig:pCp} is not connected to the hard matrix element, the left gluon propagator between the unintegrated skewed distributions $f_g(x_1,...)$ and $f_g(x_2,...)$ will be assigned a factor of $Z^2$, instead of the usual $Z$. The corresponding first order $\alpha_s$ correction therefore corresponds in part to an NLO contribution, which should not be included at lowest order in a consistent treatment.}, see the discussion in~\cite{Shuvaev:2008yn}. 

\end{itemize}

Interestingly, all of the NLO effects identified above indicate a possible increase in the rate predicted at LO, by a total factor of order $\sim 2$, although we should recall that such effects cannot be considered completely seperately from the (potentially large) variation in the cross section which can occur when different (LO and NLO) PDF sets are used.

\item[(iii)] Soft survival factors (both eikonal, $S^2_{\rm eik}$, and $S^2_{\rm enh}$). Within the Durham soft diffraction models~\cite{Ryskin:2009tj, Ryskin:2011qh, Ryskin:2009tk} their overall effect is fairly stable, fluctuating
by roughly $30-35\%$ for different sets of soft model parameters.

\item[(iv)]
For meson pair CEP, we also need to consider the impact of non--perturbative contributions, for example where the meson pair is created by double Pomeron exchange. These are only likely to be important for low invariant mass and/or transverse momentum final states, and can therefore be suppressed by imposing suitable cuts on the meson $k_\perp$. Nonetheless, for example for $\pi\pi$ CEP, where the perturbative contribution is strongly suppressed by the $J_z=0$ selection rule which operates for CEP, a cut of $k_\perp > 2$ GeV (or possibly even higher) is expected to be needed to safely suppress the predicted non--perturbative contribution. In this case, the corresponding perturbative cross sections are of order $\sim 1$ pb or lower, and so this may be difficult to access experimentally. Other observables such as $\eta(')\eta(')$ CEP are expected on the other hand to have much larger perturbative cross sections and therefore relatively smaller non--perturbative contributions. We also note that the observation of $K^+K^-$ ($p\overline{p}$, $\Lambda\overline{\Lambda}$...) CEP would also be of interest.

\end{itemize}

To summarize, if we take into account all of the main sources listed above, we estimate the uncertainty in the CEP cross section predictions (on top of the choice of PDF set) to be of order $\sim {}^{\times}_{\div}2$ for the case of $\gamma\gamma$ CEP. We would expect a larger uncertainty in the $\pi\pi$ predictions (even after the imposition of cuts to suppress the non--perturbative contributions), due to the strong suppression in the $\pi\pi$ CEP cross section from the $J_z=0$ selection rule, as in this case higher order/twist effects which allow even quite a small contribution from a $J_z=0$ component may enhance the rate. Cross section {\it ratios}, in particular $\sigma(\gamma\gamma)/\sigma(\pi\pi)$ (as well as $\sigma(\chi_c)/\sigma(\pi\pi)$ and $\sigma(\chi_b)/\sigma(\gamma\gamma)$ in the relevant mass regions), should be much less sensitive to the choice of PDFs and the survival factors, $S^2$, and should therefore provide a good test of the perturbative approach to CEP.

We have also discussed the importance of proton dissociation contributions. The standard CEP formalism assumes that the elastically scattered protons are detected in the final state. So far, however, the only experimental access to such `exclusive' events has been via large rapidity gap triggers. There is then the  issue of the importance of proton dissociation contributions to the cross section. If the experimental configuration restricts the mass of the dissociated state to a few GeV, for example by vetoing on additional particle production over a wide rapidity range, then the cross section enhancement factor is likely to be of only a few tens of $\%$. The effect of high--mass dissociation is much harder to estimate, but such events may be suppressed by restricting the transverse momentum of the central state.

We finish by recalling that all the attempts to evaluate CEP quantitatively (for sufficiently high central system mass $M_X$) are based on models which contains the same basic elements as those in the Durham model discussed in this paper. Here we consider the theoretical uncertainties contained in the various theory elements and the constraints that come from a phenomenological analysis of the existing CEP data. We have tried to make our analysis as full and complete as currently possible, although we hope in the future that further theoretical progress will be made, guided by new experimental data. It will in particular be very interesting to compare future RHIC and LHC data with our predictions for the processes discussed in this paper. For example, the forthcoming CMS data on $\gamma\gamma$ CEP should help to reduce these uncertainties and should in particular clarify the question of the most appropriate choice of PDF set~\cite{CMSgam}.

\section*{Acknowledgements}

The authors thank Mike Albrow, Erik Brucken, Yuri Dokshitzer, Wlodek Guryn, Jeff Forshaw, Jonathan Hollar, Hannes Jung, Wenbo Li, Ronan McNulty, Dermot Moran, Andy Pilkington, Risto Orava and Antoni Szczurek for useful discussions.
LHL, MGR and WJS thank the IPPP at the University of Durham for hospitality. The work by MGR was supported  by the grant RFBR 11-02-00120-a and by the Federal Program of the Russian State RSGSS-65751.2010.2. WJS acknowledges financial support in the form of an IPPP Associateship. LHL acknowledges financial support from the Cavendish Laboratory.

\bibliography{npbib}{}
\bibliographystyle{h-physrev}

\end{document}